\documentclass[prd,twoside,nofootinbib,preprintnumbers,superscriptaddress]{revtex4}

\usepackage{amsmath}
\usepackage{amssymb}  
\usepackage{mathrsfs} 
\usepackage{graphicx}
\usepackage{dcolumn}
\usepackage{color}
\usepackage{url}

\def\gsim{\raise0.3ex\hbox{$\;>$\kern-0.75em\raise-1.1ex\hbox{$\sim\;$}}}
\def\lsim{\raise0.3ex\hbox{$\;<$\kern-0.75em\raise-1.1ex\hbox{$\sim\;$}}}

\begin{document}

\preprint{IFIC/19-25}

\title{Long-lived heavy particles in neutrino mass models}

\author{Carolina Arbel\'aez R}
\email{carolina.arbelaez@usm.cl}
\affiliation{Universidad T\'ecnica Federico Santa Mar\'ia and
Centro Cient\'ifico Tecnol\'ogico de Valparaiso CCTVal
Casilla 110-V, Valparaiso, Chile}

\author{Juan Carlos Helo}
\email{jchelo@userena.cl}
\affiliation{Departamento de F\'{i}sica,
Facultad de Ciencias, Universidad de La Serena,\\
Avenida Cisternas 1200, La Serena, Chile}

\author{Martin Hirsch}
\email{mahirsch@ific.uv.es}
\affiliation{AHEP Group, Instituto de F\'{i}sica Corpuscular
- CSIC/Universitat de Val\`{e}ncia Edificio de Institutos
de Paterna, Apartado 22085, E–46071 Val\`{e}ncia, Spain}

\begin{abstract}

All extensions of the standard model that generate Majorana neutrino
masses at the electro-weak scale introduce some ``heavy'' mediators,
either fermions and/or scalars, weakly coupled to leptons. Here, by
``heavy'' we understand implicitly the mass range between a few 100
GeV up to, say, roughly 2 TeV, such that these particles can be
searched for at the LHC. We study decay widths of these mediators for
several different tree-level neutrino mass models. The models we
consider range from the simplest $d=5$ seesaw up to $d=11$ neutrino
mass models. For each of the models we identify the most interesting
parts of the parameter space, where the heavy mediator fields are
particularly long-lived and can decay with experimentally measurable
decay lengths. One has to distinguish two different scenarios,
depending on whether fermions or scalars are the lighter of the heavy
particles. For fermions we find that the decay lengths correlate with
the inverse of the overall neutrino mass scale. Thus, since no lower
limit on the lightest neutrino mass exists, nearly arbitrarily long
decay lengths can be obtained for the case where fermions are the
lighter of the heavy particles. For charged scalars, on the other
hand, there exists a maximum value for the decay length.  This maximum
value depends on the model and on the electric charge of the scalar under
consideration, but can at most be of the order of a few millimeters.
Interestingly, independent of the model, this maximum occurs always in
a region of parameter space, where leptonic and gauge boson final
states have similar branching ratios, i.e. where the observation of
lepton number violating final states from scalar decays is possible.

\end{abstract}
\maketitle

\section{Introduction}\label{sec:introduction}

The phenomenology of long-lived particles (LLPs) has received
considerable attention in the literature recently. When produced at
the Large Hadron Collider (LHC), LLPs can leave exotic signatures,
such as, for example, displaced vertices, displaced leptons, charged
tracks and others \cite{Alimena:2019zri}. All major detectors at the
LHC, ALTAS, CMS and LHCb, do search for these exotics now. In
addition, there are various experimental proposals, fully dedicated to
LLP searches, such as \texttt{MATHUSLA}
\cite{Chou:2016lxi,Curtin:2018mvb}, \texttt{CODEX-b}
\cite{Gligorov:2017nwh} and \texttt{FASER} \cite{Feng:2017uoz} or
\texttt{SHiP} \cite{Alekhin:2015byh}.  Also \texttt{MoEDAL}
\cite{Pinfold:2009oia}, originally motivated by monopole searches, now
has an LLP programme.

From the theoretical point of view there are mainly two possible
motivations for introducing LLPs
\cite{Curtin:2018mvb,Alimena:2019zri}: Dark matter and neutrino
masses. In this paper, we will focus on the latter. We will discuss
several examples of Majorana neutrino mass models and study their
predictions for the decays of the new fields that these models need to
introduce. For reasons explained below, we will concentrate on the
mass range (0.5-2) TeV, i.e. relatively ``heavy'' LLPs.

The best-known example of a Majorana neutrino mass model is the simple
type-I seesaw \cite{Minkowski:1977sc,Yanagida:1979as,Mohapatra:1979ia}.
Here, two (or more) copies of fermionic singlets, $N$, are added to
the standard model. These states, sometimes called right-handed
neutrinos, sterile neutrinos, heavy neutrinos or also ``heavy neutral
leptons'' (HNL) will mix with the SM neutrinos after electro-weak
symmetry breaking. Standard model charged and neutral current
interactions then yield a small, but non-zero production rate for
these nearly singlet states proportional to $\sum_{\alpha}|U_{\alpha i}|^2$,
where $U_{\alpha i}$ is the mixing angle between the neutrino of
flavour $\alpha=e,\mu,\tau$ and the sterile neutrino eigenstate, $N_i$. 
Recent phenomenological studies assessing the LHC sensitivity with
displaced vertices to light sterile neutrinos in various models have
been published
in~\cite{Helo:2013esa,Deppisch:2018eth,Helo:2018qej,Nemevsek:2018bbt,Lara:2018rwv,Dev:2017dui,Cottin:2018nms,Cottin:2019drg,Dercks:2018wum},
see also the review \cite{Alimena:2019zri}. The large statistics
expected at the high-luminosity LHC should allow to probe mixing
angles down to roughly $\sum_{\alpha}|U_{\alpha i}|^2\sim 10^{-9}$, 
for masses below $m_{N_1} \sim 25$ GeV \cite{Cottin:2018nms} at ATLAS
(or CMS) and below $m_{N_1} \sim 5$ GeV at \texttt{MATHUSLA} 
\cite{Helo:2018qej,Curtin:2018mvb}.

\begin{figure}
  \centering
  \includegraphics[scale=0.55]{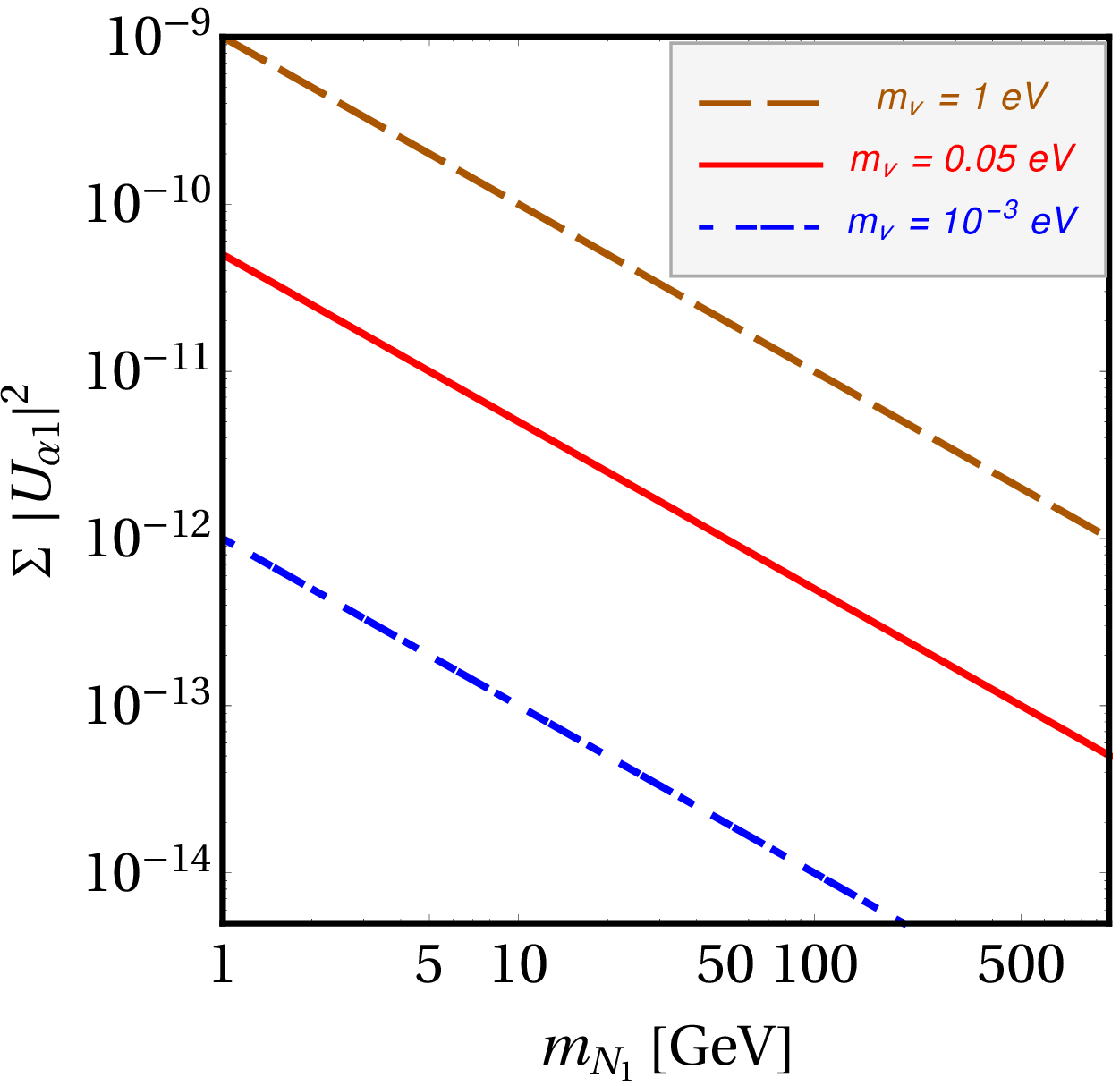}\hskip15mm
  \includegraphics[scale=0.55]{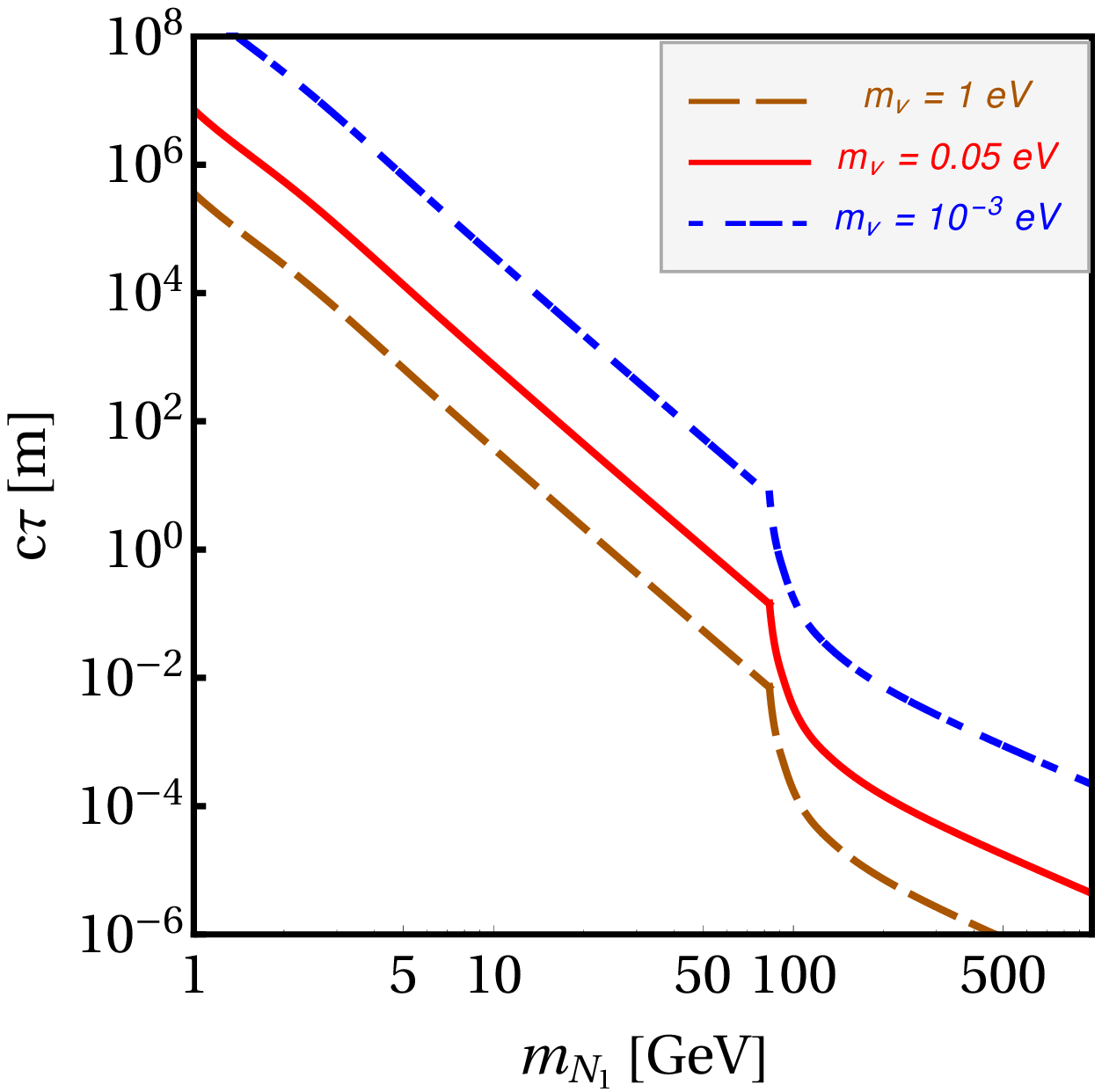}
  \caption{Naive estimate of the mixing angle (left) and decay
    length (right) of the lightest sterile neutrino in a type-I
    seesaw for a given value of the lightest active neutrino mass
    as function of the sterile neutrino mass, $m_{N_1}$.}
      \label{fig:LenSS}
\end{figure}

For the simplest seesaw type-I one can estimate an upper bound on the
decay length for a given sterile neutrino mass $m_N$, assuming that
the corresponding state contributes a certain amount to the active
neutrino mass(es). This estimate is shown in fig. (\ref{fig:LenSS}).
Here, we have used the well-known Casas-Ibarra approximation
\cite{Casas:2001sr} to reparametrize the seesaw.  For the plot we fix
the neutrino oscillation data at their best fit point (b.f.p.) values,
choose normal hierarchy and assume that the matrix ${\cal R}$ is
trivial. We then choose three different example values for the
lightest neutrino mass. Note that $m_{\nu_1}=1$ eV is experimentally
excluded, both from cosmology \cite{Aghanim:2018eyx} and from double
beta decay searches \cite{KamLAND-Zen:2016pfg,Agostini:2018tnm} and is
shown only for the sake of demonstrating the parameter dependence.
The choice of $m_{\nu_1}=0.05$ eV is motivated by the atmospheric
neutrino mass scale.  The sterile neutrino widths have been calculated
with the formulas given in \cite{Atre:2009rg,Bondarenko:2018ptm}.  The
decay lengths shown are to be understood as upper bounds, since choosing a
non-trivial ${\cal R}$ will lead in general to larger mixing and thus
to smaller decay lengths.

As fig. (\ref{fig:LenSS}) shows, for masses $m_{N_1}$ above, say
$m_W$, decay lengths drop quickly to values below $c\tau \sim 1$ mm,
except for the region of parameter space, where the lightest neutrino
mass is much smaller than $m_{\nu_1}=0.05$ eV. Comparing to
the plot on the left of the figure, however, shows that such small
values of $m_{\nu_1}$ correspond to very small values of the mixing
angle squared, $\sum_{\alpha}|U_{\alpha 1}|^2$. The production rates
of long-lived ``heavy'' steriles, i.e.  with $m_{N} \gsim 100$ GeV, is
therefore expected to be unmeasurably small in models based in seesaw
type-I.\footnote{In models with an inverse seesaw
  \cite{Mohapatra:1986bd}, larger mixings and thus, larger production
  rates are possible. However, this will lead again to {\em shorter}
  decay lengths than those shown in fig. (\ref{fig:LenSS}).}

The situation can be very different in other electro-weak scale model
of neutrino masses. The principle idea is rather simple.  Consider,
for example, the so-called seesaw type-III \cite{Foot:1988aq}. Here,
an electro-weak triplet fermion is introduced, usually denoted by
$\Sigma$. Production at the LHC can proceed, for example, via $pp \to
W^{\pm} \to \Sigma^0\Sigma^{\pm}$, i.e. with electro-weak
strength. The decays $\Sigma^0\to W^{\pm} + l_{\alpha}^{\mp}$ and
$\Sigma^{\pm} \to W^{\pm} +\nu_{\alpha}$ are, however, again mixing
suppressed \cite{Franceschini:2008pz}, as discussed above for the case
of the type-I seesaw. Thus, very small decay widths do not necessarily
imply small producion rates in general.
\footnote{Extending the standard model gauge group by, for example, an
  additional $U(1)_{B-L}$ allows to decouple production cross sections
  from decay lengths even for a type-I seesaw, as discussed in
  \cite{Jana:2018rdf,Deppisch:2019kvs}.} In this paper, we will study
this idea in some detail, considering a variety of different Majorana
neutrino models. Note that all models we consider will produce
neutrino masses at tree-level.

Majorana neutrino masses can be generated from operators of odd
dimensions, starting from $d=5$. One can write these $\Delta L=2$
operators as:
\begin{equation}\label{eq:opL2}
  {\cal O}_{\Delta L=2} = \frac{c_{\alpha\beta}}{\Lambda^{2n+1}}
    L_{\alpha}L_{\beta}HH (HH^{\dagger})^n.
\end{equation}  
Here, $c_{\alpha\beta}$ is some dimensionless coefficient and $n=0$
corresponds to the well-known Weinberg operator
\cite{Weinberg:1979sa}.  The Weinberg operator has three different
independent contractions \cite{Ma:1998dn}, which correspond - at
tree-level - to the three classical seesaws
\cite{Minkowski:1977sc,Yanagida:1979as,Mohapatra:1979ia,Schechter:1980gr,Foot:1988aq}.
Less known are the models at $d>5$. Reference \cite{Bonnet:2009ej}
presented a complete deconstruction of the $d=7$ operator at
tree-level. However, most of the ultra-violet completions (or
``models'') found in \cite{Bonnet:2009ej} require additional
symmetries, beyond those of the standard model, in order to avoid the
$d=5$ seesaw contributions.  On the other hand, there is one $d=7$
model, first presented in \cite{Babu:2009aq}, for which all $d=5$
(tree-level) contributions are {\em automatically} absent. Following
\cite{Sierra:2014rxa,Cepedello:2018rfh,Anamiati:2018cuq} we will call
such models ``genuine'' and in our numerical calculations we will
consider only genuine models. In addition to the $d=7$
model  \cite{Babu:2009aq}, called BNT model below, we will choose one
model at $d=9$ and one at $d=11$ each. These two models have been
first discussed in \cite{Anamiati:2018cuq}.

Charged particles, be they fermions or scalars, would have been seen
at LEP, if their masses are below (80-115) GeV
\cite{Tanabashi:2018oca}, with the exact number depending on the decay
modes and nature of the particle under consideration.  Thus, none of
the models we consider in this paper can work with an overall mass
scale below roughly 100 GeV. Also the LHC will provide important
constraints on all our models. Here, however, the situation is much
more complicated. Consider for example the seesaw type-II
\cite{Schechter:1980gr}. Here, a scalar triplet is added to the
standard model and ATLAS has searched for this state in $pp\to
\Delta^{++}\Delta^{--} \to 4 \ell$ \cite{Aaboud:2017qph}.  The limits
are in the range (770 - 870) GeV, depending on the lepton flavour,
assuming Br($\Delta^{\pm\pm}\to\ell^{\pm}_{\alpha}\ell^{\pm}_{\beta}$)
equal to 1 (for $\alpha,\beta=e,\mu$).  CMS \cite{CMS:2017pet} has
studied both pair- and associated production of scalars and obtained
limits for leptonic final states similar to the ones by ATLAS.  ATLAS
has also considered $pp\to \Delta^{++}\Delta^{--} \to 4 W$. However,
lower limits for a $\Delta^{++}$ decaying to gauge bosons are
currently only of order 220 GeV \cite{Aaboud:2018qcu}.  For the
fermions of seesaw type-III CMS quotes a lower limit of $\sim 840$ GeV
\cite{Sirunyan:2017qkz}, assuming a ``flavour democratic'' decay,
i.e. equal branching ratios to electrons, muons and taus. Weaker
limits would be obtained for states decaying mostly to taus. Note that
this search \cite{Sirunyan:2017qkz} assumes that all decays are
prompt. For the more complicated final states of our $d>5$ models no
dedicated LHC searches exist so far, but one can expect that current
sensitivities should be somewhere between ($0.5-1$) TeV, depending on
whether one considers scalars or fermions.  The high-luminosity LHC
will of course be able to considerably extend this mass range. Thus,
we will consider (roughly) the mass range (0.5-2) TeV in this paper.

The rest of this paper is organized as follows. We will give the
particle content and the Lagrangians of the models used in the
numerical parts of this paper in section \ref{sec:models}. We
divide then the numerical discussion into two sections. First,
in section \ref{sec:fermionicdesintegration} we calculate the
decays of exotic fermions, while section \ref{sec:scalardesintegrations}
discusses the decays of exotic scalars. In both sections, the
main emphasis is on identifying the regions in parameter space,
where the decay widths of the lightest exotic particles are
sufficiently small that experiments will be able to find displaced
vertices or charged tracks, depending on whether the particles
considered are neutral or charged. We then close with a short
summary and discussion. 

\section{Models} \label{sec:models}

In this section we will discuss the basics of the different neutrino
mass models that we consider in the numerical part of this work. All
models studied here generate neutrino masses at tree-level. The lowest
order operator to generate Majorana neutrino masses is the Weinberg
\cite{Weinberg:1979sa} operator at $d=5$, which at tree-level can be
realized in three different ways. For higher dimensional operators
generating neutrino masses, we will consider one example model each at
$d=7$, $d=9$ and $d=11$. The models we selected are all ``genuine''
models in the sense that they give automatically the leading
contribution to the neutrino mass without resorting to any additional
symmetry beyond the SM ones.

To fix notation, with the exception of the classical three seesaw
mediators (see below), we will use ${\bf R}_Y$ for $SU(2)_L$
representations with hypercharge $Y$ and add a superscript $F$ or $S$
to denote fermions and or scalars. For example, ${\bf 3}^F_1$ is a
fermionic triplet with hypercharge one.  The different components of
these representations are then written classified by their electric
charge. For example, ${\bf 3}^F_1$ is written in components as
($F^{++}_3,F^{+}_3,F^{0}_3$), where we exchange ${\bf R}$ and $F$ to
avoid confusion between multiplets and charge eigenstates.

\subsection{$d=5$}

\begin{figure}
  \centering
  \includegraphics[scale=0.8]{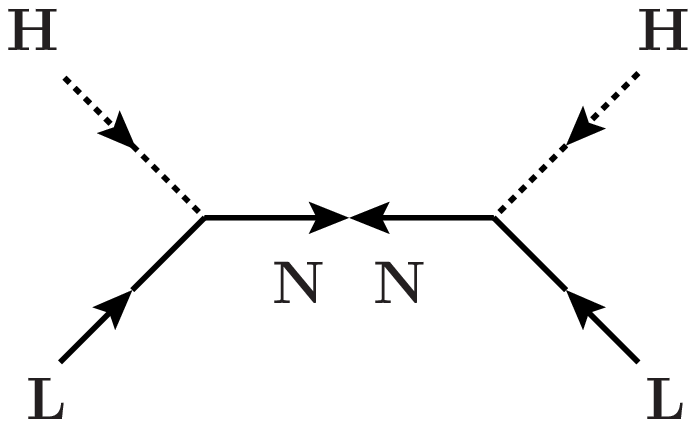}\hskip5mm
  \includegraphics[scale=0.8]{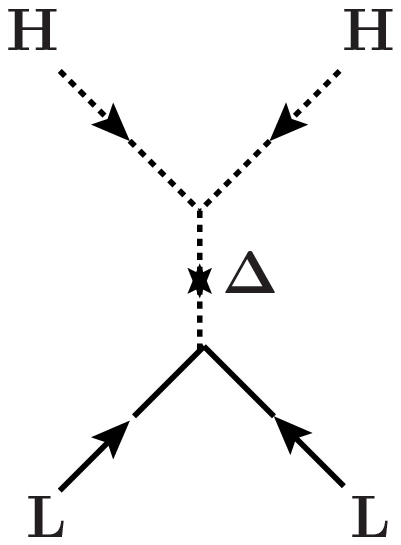}\hskip5mm
  \includegraphics[scale=0.8]{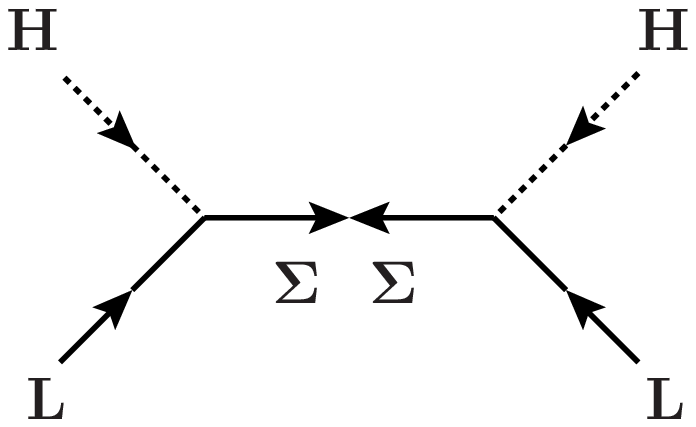}
  \caption{Tree-level neutrino mass diagrams for the $d=5$ Weinberg
    operator. These are the three well-known variants of the seesaw,
    from left to right: Type-I, -II and -III.}
      \label{fig:d5diags}
\end{figure}

At $d=5$ there are three models that generate neutrino masses at
tree-level. These are known as seesaw type-I, type-II and type-III in
the literature. Seesaw type-I adds fermionic singlets to
the standard model (SM), ${\bf 1}^F_0 \equiv N$. Type-II adds a
scalar triplet to the SM, ${\bf 3}^S_1 \equiv \Delta$, while type-III
uses fermionic triplets, ${\bf 3}^F_0 \equiv \Sigma$. For these three
fields the names $N$, $\Delta$ and $\Sigma$ are customary in the
literature; we follow this convention. The neutrino mass diagrams
for the three classical seesaws are shown in fig. \ref{fig:d5diags}.

These three models are very well-known and we give the Lagrangians
here only to fix the notation.  The Lagrangian for type-I seesaw is
given by:
\begin{equation}\label{eq:sst1}
  {\cal L} = {\cal L}^{\rm SM} + Y_{\nu} {\overline L} H^{\dagger} N
             +  \frac{1}{2} M_N {\overline N^c} N .
\end{equation}
Here, $M_N$ is a matrix with eigenvalues $m_{N_i}$.
\footnote{We use capital letters for matrices, lowercase letters for
  eigenvalues.}
For the seesaw type-II one has:
\begin{equation}\label{eq:sst2}
  {\cal L} = {\cal L}^{\rm SM} + Y_{\Delta} {\overline L^c} \Delta L 
           + \mu_{\Delta} H \Delta^{\dagger} H + m_{\Delta}^2 |\Delta|^2 \cdots
\end{equation}
Here, we neglected to write down other terms in the scalar potential
which are not directly relevant for neutrino mass generation. Recall,
that the term proportional to $\mu_{\Delta}$ induces a vacuum
expectation value for the triplet: $v_{\Delta} \simeq
\mu_{\Delta}\frac{v^2}{m_{\Delta}^2}$. Here, $v$ is the standard model
Higgs vev. Finally, for type-III we write:
\begin{equation}\label{eq:sst3}
  {\cal L} = {\cal L}^{\rm SM} + Y_{\Sigma} {\overline L} H^{\dagger}\Sigma
             + \frac{1}{2} M_{\Sigma}\Sigma\Sigma .
\end{equation}
In these equations we have suppressed generation indices. Recall,
however, that for type-I and type-III seesaw there is one non-zero
active neutrino mass per generation of singlet/triplet field added.
Thus, for these cases at least two copies of extra fermions are needed.
In our numerical studies we always use three copies of extra fermions.
All matrices $Y_{\nu}$, $Y_{\Delta}$, $Y_{\Sigma}$ and $M_N$ and
$M_{\Sigma}$ are then ($3,3$) matrices. 

For setting up the notation, here we give also the mass matrices
for the neutral and charged fermions in seesaw type-III. The
neutral fermion mass matrix, in the basis ($\nu,\Sigma^0$),
is given by
\begin{equation}\label{eq:mzeroIII}
  M^0 =
  \left(
  \begin{array}{cc}
   0& \frac{v}{\sqrt{2}}Y_{\Sigma} \\ 
   \frac{v}{\sqrt{2}}Y_{\Sigma}^T &M_{\Sigma}
  \end{array}
\right),
\end{equation}
while the charged fermion mass matrix, in the basis
($l,\Sigma^-$), is:
\begin{equation}\label{eq:mpIII}
  M^+ =
  \left(
  \begin{array}{cc}
   \frac{v}{\sqrt{2}}Y_{l}& \frac{v}{\sqrt{2}}Y_{\Sigma} \\ 
   0 &M_{\Sigma}
  \end{array}
\right).
\end{equation}
Note that the mixing between heavy and light states in both,
the neutral and the charged sectors, is to first approximation 
given by $(Y_{\Sigma}v).M_{\Sigma}^{-1}$.

\subsection{$d=7$}

\begin{figure}
  \centering
  \includegraphics[scale=0.5]{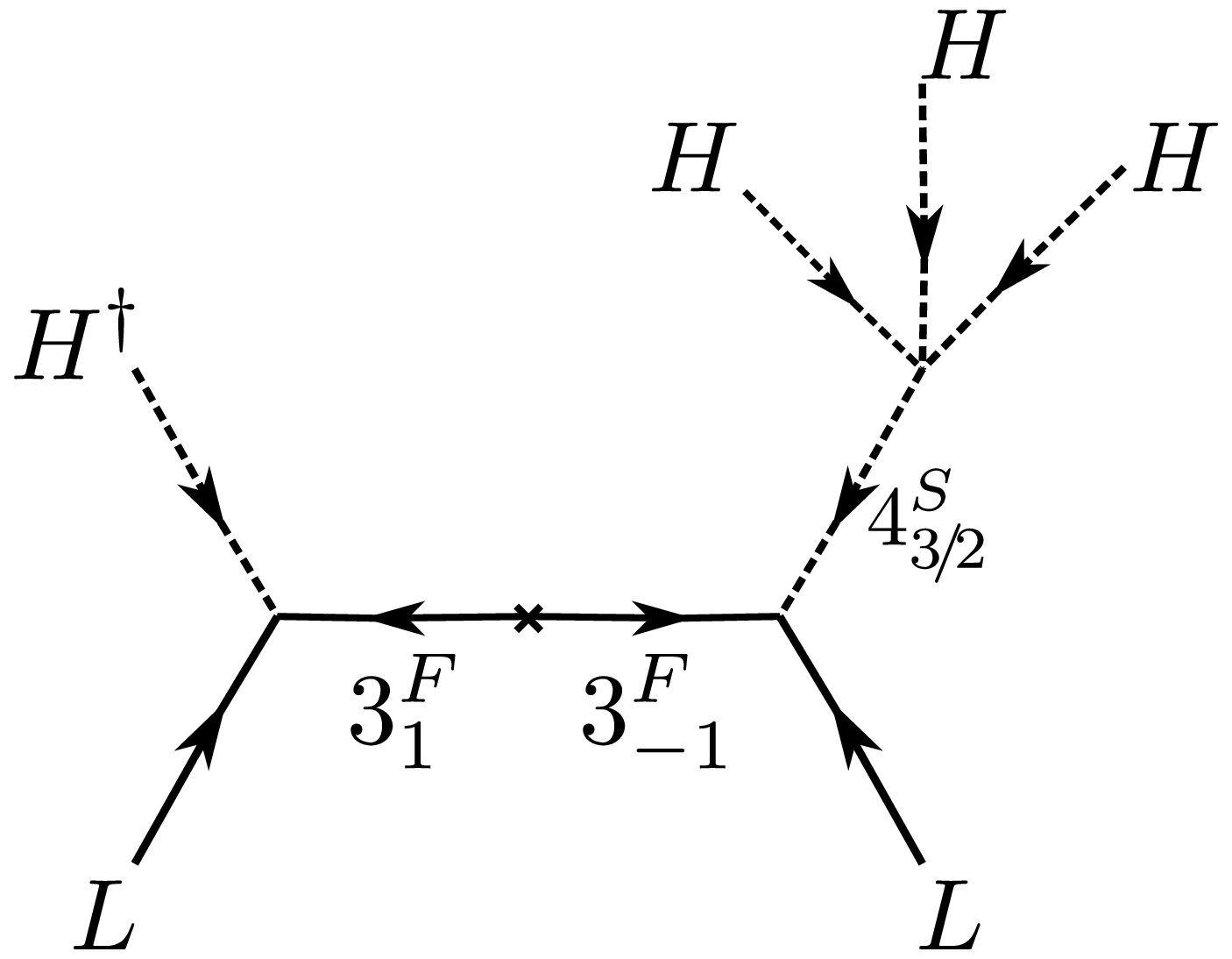}
  \caption{Tree-level neutrino mass diagram for the genuine $d=7$
    model~\cite{Babu:2009aq}.}
      \label{fig:d7diag}
\end{figure}

The model we consider at $d=7$ first appeared in reference
~\cite{Babu:2009aq}; we therefore call it the BNT model.  The
phenomenology of this model has been studied also in
\cite{Ghosh:2017jbw,Ghosh:2018drw}.  In the BNT model, two kinds of
fields are added to the SM; a scalar quadruplet, ${\bf 4}^S_{3/2}$ and
a fermionic triplet, ${\bf 3}^F_1$, together with its vector partner
${\bf 3}^F_{-1}$.  The Lagrangian of this model contains the terms
\begin{equation}\label{eq:BNT}
  {\cal L} = {\cal L}^{\rm SM}
            + Y_{\Psi} L H^{\dagger} {\bf 3}^F_1
            + Y_{\overline \Psi} {\bf 4}^S_{3/2} {\bf 3}^F_{-1}L
            + M_{\bf 3} {\bf 3}^F_1 {\bf 3}^F_{-1}
            + m_4^2 |{\bf 4}^S_{3/2}|^2
            + \lambda_5 H H H ({\bf 4}^S_{3/2})^{\dagger} + \cdots,
\end{equation}
Here, we have again written only those terms that appear in the
diagram for neutrino masses, see fig. (\ref{fig:d7diag}). The last
term in eq. (\ref{eq:BNT}) induces a vacuum expectation value (vev)
for the neutral component of the ${\bf 4}^S_{3/2}$, even for $ m_4^2
\ge 0$.  Roughly one can estimate
\begin{equation}\label{eq:v4}
v_{\bf 4} \simeq - \lambda_5 \frac{v^3}{m_4^2}.
\end{equation}
The SM $\rho$ parameter \cite{Tanabashi:2018oca} requires that this
vev is small, we estimate $v_{\bf 4} \lsim 3.5$ GeV.

In the calculation of the neutrino mass matrix, one can use eq.
(\ref{eq:v4}) to rewrite the full neutral fermion mass matrix, in
the basis ($\nu,F^0_3,{\overline F}^0_{3}$):\footnote{${\overline F}^0_{3}$
  denotes the neutral component in ${\bf 3}^F_{-1}$.}
\begin{equation}\label{eq:mzeroBNT}
  M^0 =
  \left(
  \begin{array}{ccc}
   0& \frac{v}{\sqrt{2}}Y_{\Psi}& \frac{v_{\bf 4}}{\sqrt{2}}Y_{\overline \Psi}^T\\
   \frac{v}{\sqrt{2}}Y_{\Psi}^T & 0 &M_{\bf 3}\\
   \frac{v_{\bf 4}}{\sqrt{2}}Y_{\overline \Psi} &M_{\bf 3}^T& 0
  \end{array}
\right).
\end{equation}
This matrix has the same structure as found in the so-called ``linear
seesaw'' models \cite{Akhmedov:1995ip,Akhmedov:1995vm}. The light
neutrino masses are therefore given approximately as
\begin{equation}\label{eq:mnuBNT}
  m_{\nu} \simeq \frac{1}{2} 
  \left(Y_{\Psi}M_{\bf 3}^{-1}Y_{\overline \Psi}^T+
                      Y_{\overline\Psi}M_{\bf 3}^{-1}Y_{\Psi}^T\right) v v_{\bf 4}.
\end{equation}
Note the additional suppression of the neutrino masses by a factor
($v_{\bf 4}/v$) compared to the standard seesaw. 

The charged fermion mass matrix in the basis $(l, F_{3}^-)$, is
given by:
\begin{equation}
  M^{+}=  \left(
\begin{array}{c c}
\frac{v Y_{e}}{\sqrt{2}} & -\frac{v}{2}Y_{\Psi} \\
0 & M_{\bf 3}
\end{array}
\right)
\label{eq:mpBNT}
\end{equation}
Note that the neutrino masses require $Y_{\Psi}$ and/or
$Y_{\overline\Psi}$ to be small, see eq. (\ref{eq:mnuBNT}). Thus,
unless there is a large hierarchy imposed on the couplings $Y_{\Psi}$
and $Y_{\overline\Psi}$, one expects the mixing between the heavy
charged states and the charged leptons to be small.

\subsection{$d=9$}

\begin{figure}
  \centering
  \includegraphics[scale=0.4]{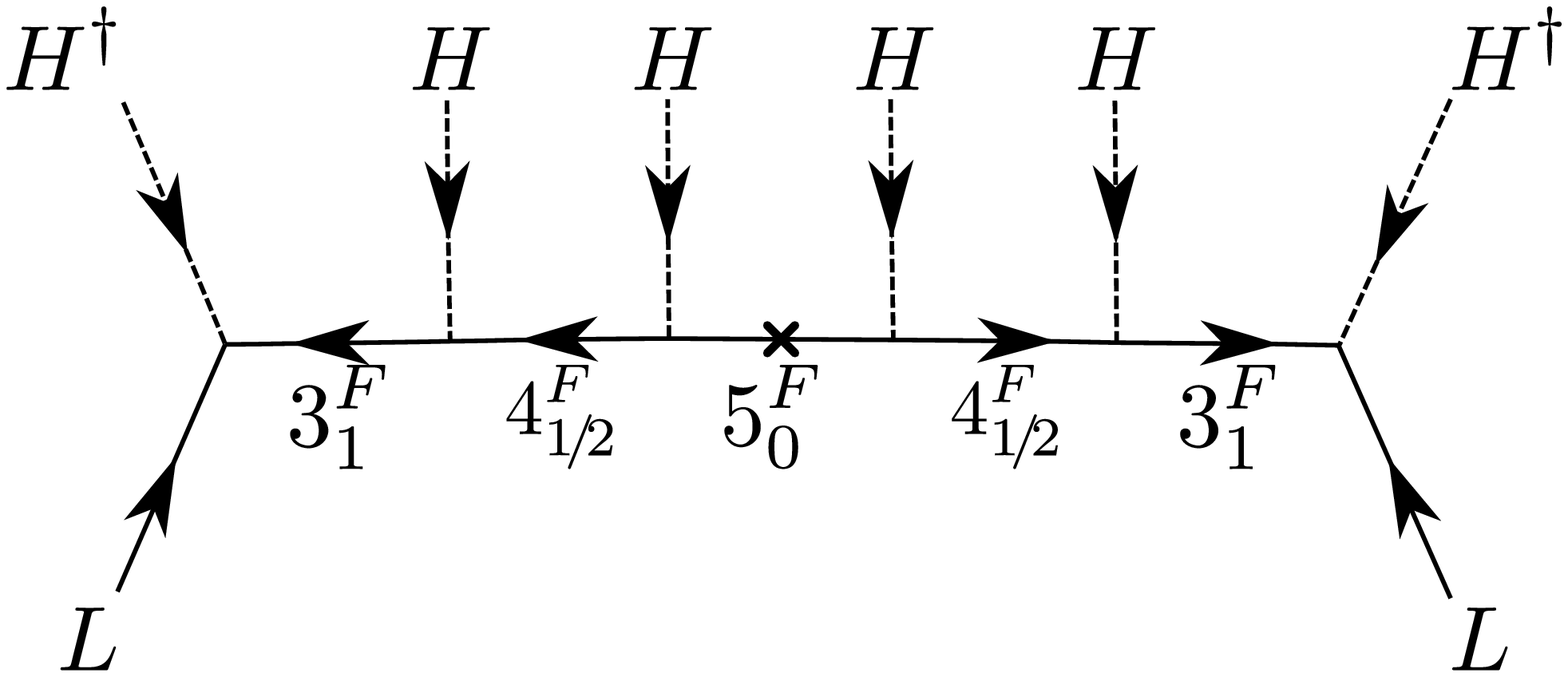}\hskip10mm
  \includegraphics[scale=0.13]{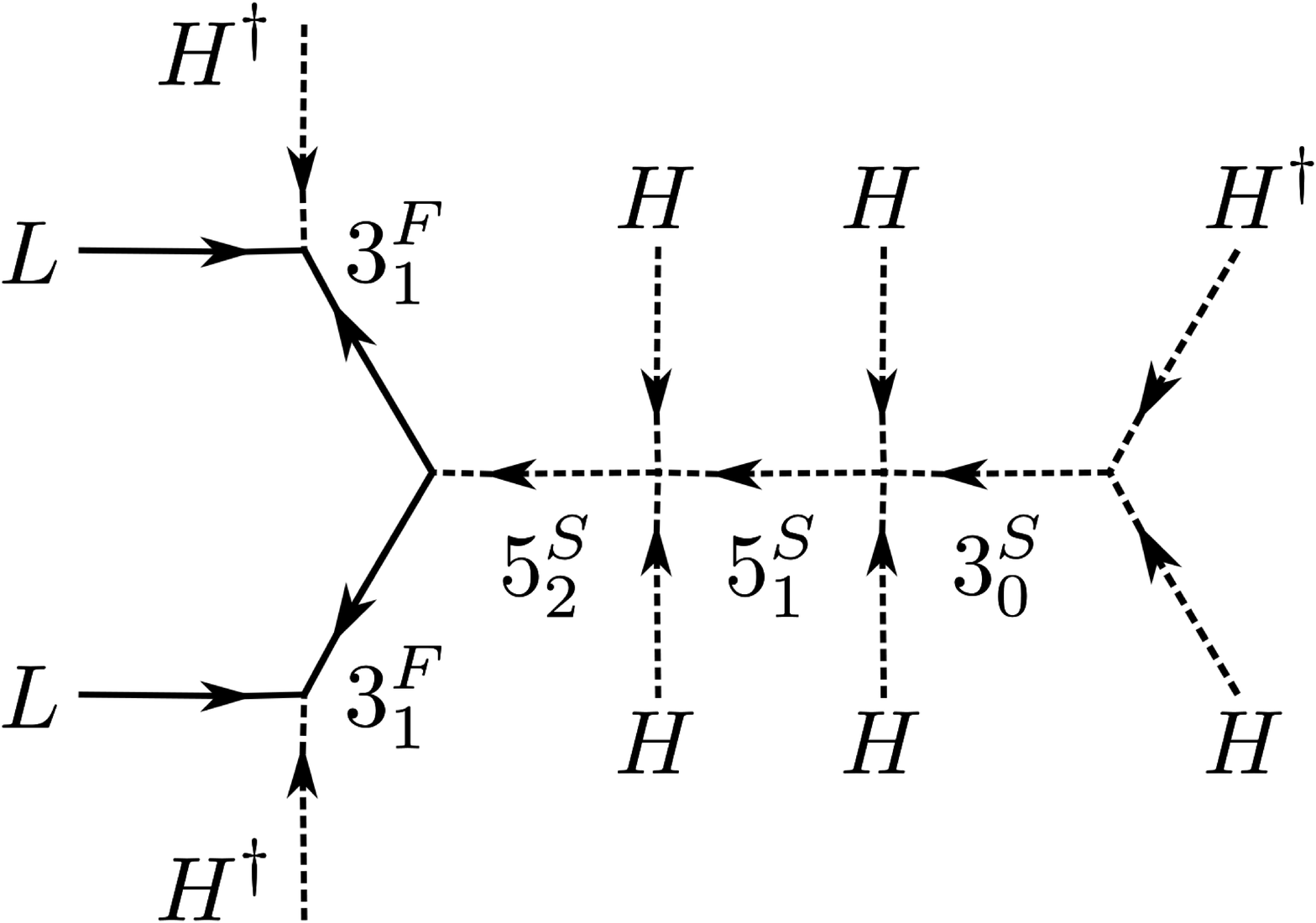}
  \caption{Tree-level neutrino masses diagrams for two different
    higher-dimensional neutrino mass models. A $d=9$ diagram to the
    left and a $d=11$ diagram to the right.}
  \label{fig:d9d11}
\end{figure}

As shown in \cite{Anamiati:2018cuq}, there are two genuine tree-level
neutrino mass models at $d=9$. model-I is the simpler variant, since
it contains only three new fermions, ${\bf 3}_1^F$, ${\bf 4}_{1/2}^F$
and ${\bf 5}_0^F$, together with their vector partners.  The
Lagrangian of the model can be written as:
\begin{equation}
  \label{eq:lagd9m1}
        {\cal L} = {\cal L}_{SM}+ {\cal L}_{Yuk} + {\cal L}_{mass},
\end{equation}
where
\begin{equation}
  \begin{aligned}
    \label{eq:yuk9m1}
          {\cal L}_{Yuk} &= Y_{\Psi} \, L  {\bf 3}_{1}^F H^{\dagger} 
          +  {\overline Y}_{\bf 34} \, {\bf 3}_{-1}^F   {\bf 4}_{1/2}^F H
          +  Y_{\bf 34} \, {\bf 3}_{1}^F  {\bf 4}_{-1/2}^F  H^{\dagger}
          \\ 
          & \quad +
        {\overline Y}_{\bf 45} \, {\bf 4}_{1/2}^F  {\bf 5}_{0}^F  H^{\dagger}
          +   Y_{\bf 45}  \, {\bf 4}_{-1/2}^F {\bf 5}_{0}^F   H
  \end{aligned}
\end{equation}
and
\begin{equation}
  \label{eq:mass9m1}
        {\cal L}_{mass} = M_{\bf 3} {\bf 3}_{1}^F{\bf 3}_{-1}^F
        + M_{\bf 4} {\bf 4}_{1/2}^F{\bf 4}_{-1/2}^F
        + M_{\bf 5} {\bf 5}_{0}^F{\bf 5}_{0}^F .
\end{equation}
The mass matrix for the neutral states, in the basis
($\nu,F^0_3,{\overline F}^0_3,F^0_4,{\overline F}^0_4,F^0_5$),
is given by
\begin{equation}\label{eq:mnuDim9}
  M^0 =
  \left(
  \begin{array}{cccccc}
    0 & \frac{v}{\sqrt{2}}Y_{\Psi} & 0 & 0 & 0 & 0\\
    \frac{v}{\sqrt{2}}Y_{\Psi}^T & 0 &M_{\bf 3} & 0 &
    \frac{v}{\sqrt{6}}Y_{\bf 34} & 0\\
   0 &M_{\bf 3}^T& 0 & \frac{v}{\sqrt{6}}{\overline Y}_{\bf 34} & 0 & 0 \\
   0 & 0 & \frac{v}{\sqrt{6}}{\overline Y}_{\bf 34}^T & 0 & M_{\bf 4}
   &  \frac{v}{2}{\overline Y}_{\bf 45} \\
   0 &  \frac{v}{\sqrt{6}}Y_{\bf 34}^T & 0 & M_{\bf 4}^T & 0
   &\frac{v}{2} Y_{\bf 45} \\
   0 & 0 & 0 &  \frac{v}{2}{\overline Y}_{\bf 45}^T
   & \frac{v}{2} Y_{\bf 45}^T & M_{\bf 5} \\
  \end{array}
\right).
\end{equation}
This matrix has an iterated seesaw structure, as can be understood
also from the neutrino mass diagram shown in fig. (\ref{fig:d9d11}),
to the left. Repeatedly applying the seesaw approximation, the light
neutrino mass matrix can be roughly estimated as:
\begin{equation}
  \label{eq:numass1}
  m_{\nu} \sim \frac{1}{48}
  Y_{\Psi}M_{\bf 3}^{-1}{\overline Y}_{\bf 34} M_{\bf 4}^{-1}Y_{\bf 45}
  M_{\bf 5}^{-1}Y_{\bf 45}^T(M_{\bf 4}^T)^{-1}{\overline Y}_{\bf 34}^T
  (M_{\bf 3}^T)^{-1}Y_{\Psi}^T v_{SM}^6 + \cdots
\end{equation}
Here, we have neglected terms proportional to $Y_{\bf 34}$ and
${\overline Y}_{\bf 45}$, since they are formally of higher order
(i.e. $d=11,13$ contributions to $m_{\nu}$). Note that, for generating
non-zero neutrino masses all matrices $Y_{\Psi}$, 
${\overline Y}_{\bf 34}$ and $Y_{\bf 45}$ must contain non-zero
entries, while non-zero terms for $Y_{\bf 34}$ and ${\overline Y}_{\bf
  45}$ are not required.

Finally, for this model, the charged fermion matrix, in the basis
$(l, F_{3}^{-}, F_{4}^{-}, (F_{4}^{+})^c, F_{5}^{-})$, is given by:
\begin{equation}
    M^{+}=\begin{pmatrix}
    \frac{v}{\sqrt{2}}Y_{e}&-\frac{v}{2}Y_{\Psi}&0&0&0\\
    0&M_{\bf 3}&\frac{1}{\sqrt{3}}v{\overline Y}_{\bf 34}&0&0\\
    0&\frac{1}{\sqrt{3}}v{Y}_{\bf 34}^T&0&M_{\bf 4}
    &-\frac{v}{2\sqrt{2}}{\overline Y}_{\bf 45}\\
    0&0&M_{\bf 4}&0&\frac{\sqrt{3}}{2\sqrt{2}}v Y_{\bf 45 }\\
    0&0&-\frac{\sqrt{3}}{2\sqrt{2}}v {\overline Y}_{\bf 45 }^T
    & \frac{v}{2\sqrt{2}}Y_{\bf 45}^T&-M_{\bf 5}
    \end{pmatrix} .
    \label{eq:dim9pmatrix}
\end{equation}

\subsection{$d=11$}

Also at $d=11$ level there are two genuine neutrino mass models at
tree-level \cite{Anamiati:2018cuq}. The neutrino mass diagram of
model-I is shown in fig. \ref{fig:d9d11}, to the right.  Again, we
have chosen this model over model-II since it has the smaller
particle content. Model-I needs one exotic fermion, ${\bf 3}^F_1$
(plus vector partner) and three different scalars: ${\bf 5}^S_2$,
${\bf 5}^S_1$ and ${\bf 3}^S_0$, Here we write down only the
Lagrangian terms important for generating neutrino masses:
\begin{equation}
  \begin{aligned}
    \label{eq:d11Lag}
          {\cal L} & \propto \lambda_{55}({\bf 5}_{2}^S)^{\dagger}{\bf 5}_1^S H H
          + \lambda_{53}({\bf 5}_{1}^S)^{\dagger}{\bf 3}_0^S H H
          + \mu_3 {\bf 3}_0^S H H^{\dagger}
          + m_3^2 |{\bf 3}_0^S|^2 
          \\ 
          & \quad  + m_{5_1}^2 |{\bf 5}_{1}^S|^2
          + m_{5_2}^2 |{\bf 5}_{2}^S|^2
          + Y_{\Psi} L  {\bf 3}_{1}^F  H^{\dagger}
          + Y_{3} {\bf 3}_{1}^F  {\bf 3}_{1}^F ({\bf 5}_2^S)^{\dagger}
          + Y_{\overline{3}} {\bf 3}_{-1}^F  {\bf 3}_{-1}^F {\bf 5}_2^S
          + M_{\bf 3} {\bf 3}_{-1}^F{\bf 3}_{1}^F.
  \end{aligned}
\end{equation}
The neutral components of the scalars ${\bf 3}_0^S$, ${\bf 5}_{1}^S$
and ${\bf 5}_{2}^S$ will acquire vevs, $v_3$, $v_{51}$ and $v_{52}$
respectively. Solving the tadpole equations gives:
\begin{eqnarray}\label{eq:tadv3}
  v_3 & = & \frac{\mu_3 v^2 (16 m_{5_1}^2 m_{5_2}^2 - \lambda_{55}^2 v^4)}
  {(-32 m_3^2 m_{5_1}^2 m_{5_2}^2 + 2 \lambda_{55}^2 m_3^2 v^4
    + 4 \lambda_{53}^2 m_{5_2}^2 v^4)}
  \sim - \frac{\mu_3 v^2}{2 m_3^2} , \\ \label{eq:tadv51}
  v_{51} &=& - \frac{2 \sqrt{2} \lambda_{53} m_{5_2}^2 \mu_33 v^4}
  {(-16 m_3^2 m_{5_1}^2 m_{5_2}^2 + \lambda_{55}^2 m_3^2 v^4 + 2 \lambda_{53}^2
    m_{5_2}^2 v^4)}
  \sim  \frac{\lambda_{53} \mu_3 v^4}{4\sqrt{2} m_3^2 m_{5_1}^2} ,
  \\ \label{eq:tadv52}
  v_{52} & = & -\frac{\lambda_{53} \lambda_{55} \mu_3 v^6}
  {\sqrt{2} (-16 m_3^2 m_{5_1}^2 m_{5_2}^2 + \lambda_{55}^2 m_3^2 v^4 + 
    2 \lambda_{53}^2 m_{5_2}^2 v^4)}
  \sim
 \frac{\lambda_{53} \lambda_{55} \mu_3 v^6}{16\sqrt{2}m_3^2 m_{5_1}^2 m_{5_2}^2}.
\end{eqnarray}
Again, all these vevs have to be numerically small, due to the precise
measurements for the $\rho$ parameter \cite{Tanabashi:2018oca}.
Eq. (\ref{eq:tadv52}) allows to trade the parameters of the scalar
sector for $v_{52}$ in the calculation of the neutral fermion mass
matrix. The resulting mass matrix, in the basis ($\nu,F^0_3,{\overline
  F}^0_{3}$), is given by:
\begin{equation}\label{eq:mnuD11}
  M^0 =
  \left(
  \begin{array}{ccc}
   0& \frac{v}{\sqrt{2}}Y_{\Psi}& 0\\
   \frac{v}{\sqrt{2}}Y_{\Psi}^T & \frac{1}{\sqrt{2}} Y_{3} v_{52} &M_{\bf 3}\\
   0 &M_{\bf 3}^T& \frac{1}{\sqrt{2}} Y_{\overline{3}} v_{52}
  \end{array}
\right).
\end{equation}
This matrix is of the same form as the one found in ``inverse seesaw''
models \cite{Mohapatra:1986bd}.  A simple estimate for the neutrino
mass from the diagram on the right in fig.~\ref{fig:d9d11} gives:
\begin{equation}
  \label{eq:numass1-11}
  m_{\nu} \simeq \frac{v_{52}v^2}{2\sqrt{2}} Y_{\Psi} M_{\bf 3}^{-1} Y_{\overline{3}}
  (M^T_{\bf 3})^{-1}Y_{\Psi}^T . 
\end{equation}

\bigskip
We have implemented all models discussed in this section in
\texttt{SARAH} \cite{Staub:2012pb,Staub:2013tta}. In all cases, we use
three generation of the new fermion fields, although we mention that
neutrino oscillation fits in principle could be done with less copies
of fields. These implementations allow to generate automatically
\texttt{SPheno} routines \cite{Porod:2003um,Porod:2011nf} for the
numerical evaluation of mass spectra, mixing matrices, 2-body and
fermionic 3-body decays. \texttt{SARAH} also allows to generate model
files for \texttt{MadGraph}
\cite{Alwall:2007st,Alwall:2011uj,Alwall:2014hca}, which we use for
the numerical calculation of several multi-particle final state decay
widths, see section \ref{sec:scalardesintegrations}.  Finally, we use
the package \texttt{SSP} \cite{Staub:2011dp} to perform parameter
scans.

\section{Fermionic decays} \label{sec:fermionicdesintegration}

In this section, we will study the decay of the heavy fermionic fields
that mediate the $d=5$, $d=7$ and $d=9$ tree-level neutrino mass
generation mechanisms described in the last section. For each model,
we identify the most interesting part of the parameter space where the
heavy mediator field has an experimentally measurable decay length.
In this section, we always assume that the exotic scalars of the
different models are heavier than the fermions, such that they do not
appear in the decay chains.

The CMS search for the fermions of the seesaw type-III
\cite{Sirunyan:2017qkz}, will give limits on all models we are
discussing in this section too, as long as the decays of the fermions
are prompt. The limits derived in \cite{Sirunyan:2017qkz}, depend
heavily on the lepton flavour the heavy fermion decays to.
Varying the branching ratios to $e,\mu,\tau$ between zero and
one, limits in the range (390-930) GeV were found. The lower
end of this range corresponds to fermions decaying to taus,
while the upper end corresponds to fermions decaying 100 \% to
muons. We expect that limits on the fermionic states of the other
models we are considering should lie in a similar range. However,
we stress that here we are mostly interested in long-lived fermions,
so the search \cite{Sirunyan:2017qkz} is not directly applicable.

Let us start the discussion with some general observations. In any of
the different models, the heavy fermions $F$ will decay to a light
fermion $f$ plus a boson $V$ as $F\rightarrow f+V$.\footnote{Since we
  consider more than one generation of heavy fermions, there will also
  be decays, such as $F_i \to F_j + V$ for the heavier fermionic
  states. In the numerical calculation of the total widths of these
  fermions, all decay channels are taken into account automatically by
  SPheno, but we will not discuss these in details here.}  Here, $V$
can be either a $W$, a $Z^0$ or a Higgs boson, $h$. Consider, for
example, the charged-current vertex. In the limit where the heavy
fermion mass is much larger than the $W$ mass the decay width
for $\Gamma(F^0\rightarrow f^-+W^+)$ is described approximately by:
\begin{equation}
  \Gamma(F^0_i \rightarrow W^+ + l^-_j)
  \sim \frac{1}{32\pi}|V_{ij}|^2\frac{m_{F}^3}{m_{W}^2},
  \quad     V_{ij}=\frac{g}{\sqrt{2}}U_{\nu_{i \alpha}}U^{\dagger}_{l_{\alpha j}}
\label{eq:dwII}
\end{equation}
Here, $U_{\nu}$ diagonalizes the mass matrix of the neutral states,
and $U_{l}$ diagonalizes the mass matrix of the charged states. Both,
in the SM and in seesaw type-I, we can choose a basis where $U_{l}$ is
diagonal, but in more general models this is not the case.
Off-diagonal entries, connecting heavy and light states, will be
suppressed in both, $U_{\nu}$ and $U_{l}$, with the suppression
usually of the same order in both matrices. Typically one finds
$|V_{ij}|^2 \sim \frac{m_{\nu}}{m_{F}}$, for heavy-to-light
transitions.  Similar suppression factors are found for final states
involving $Z$ and $h$ bosons. This explains the smallness of the heavy
particle widths qualitatively.  In our numerical analysis, however, we
always use \texttt{SPheno} \cite{Porod:2003um,Porod:2011nf}, which
diagonalizes all mass matrices exactly.

\subsection{$d=5$}

In seesaw models, both, $U_{l}$ and $U_{\nu}$, are in general larger
matrices than $(3\times3)$. In seesaw type-I $U_{l}$ and $U_{\nu}$ are
$(3\times3)$ and $(6\times6)$ matrices respectively. For an estimation
of the decay length for a sterile neutrino in seesaw type-I see the
introduction.  Here, we will discuss the phenomenologically more
interesting case of the seesaw type-III.

For seesaw type-III, both $U_{\nu}$ and $U_{l}$ are $(6\times6)$
matrices.  We can express $Y_{\Sigma}$ in terms of the measured
neutrino masses and angles, using the Casas-Ibarra parametrization
\cite{Casas:2001sr}:
\begin{equation}\label{eq:CI}
  Y_{\Sigma} = \frac{i}{v}\sqrt{\hat M_{\Sigma}^{-1}}{\cal R}\sqrt{{\hat m}_{\nu}}
  U_{\nu}^{\dagger}.
\end{equation}  
Here, as usual, ${\hat M_{\Sigma}}$ is diagonal, ${\cal R}$ is an
arbitrary orthogonal matrix, ${\hat m}_{\nu}$ are the light neutrino
mass eigenvalues and $U_{\nu}$ is assumed to be the mixing matrix, as
measured in oscillation experiments. Note, that since $U_l$ is not
diagonal in general, the latter is an approximation. However, mixing
between heavy and light sectors are of order
$(Y_{\Sigma}v).M_{\Sigma}^{-1}$, for both neutral and charged
fermions, see section \ref{sec:models}.  Corrections to
eq. (\ref{eq:CI}) are therefore usually small. We note that this is
particularly so in the region of parameter space, in which we are
interested in: Large decay lengths require small values of
$(Y_{\Sigma}v).M_{\Sigma}^{-1}$. We thus put in the numerical scans
${\cal R}=1$. Smaller decay lengths than the ones shown below could be
obtained, for a matrix ${\cal R}$ with complex angles with
large absolute values $|z|^2\gg 1$. 
\footnote{A general ($3,3$) orthogonal matrix can be written as a
  product of three complex rotation angles $z_i$.  Following
  \cite{Anamiati:2016uxp} we can write $\zeta_i = \kappa_i \cdot
  e^{2i\pi\, x_i}$, such that $\sin(z_i)^2+\cos(z_i)^2=1$, while
  $|\kappa_i|^2$ can be $|\kappa_i|^2>1$.}

\begin{figure}
\begin{center}
\begin{tabular}{cc}
\includegraphics[width=0.5\textwidth]{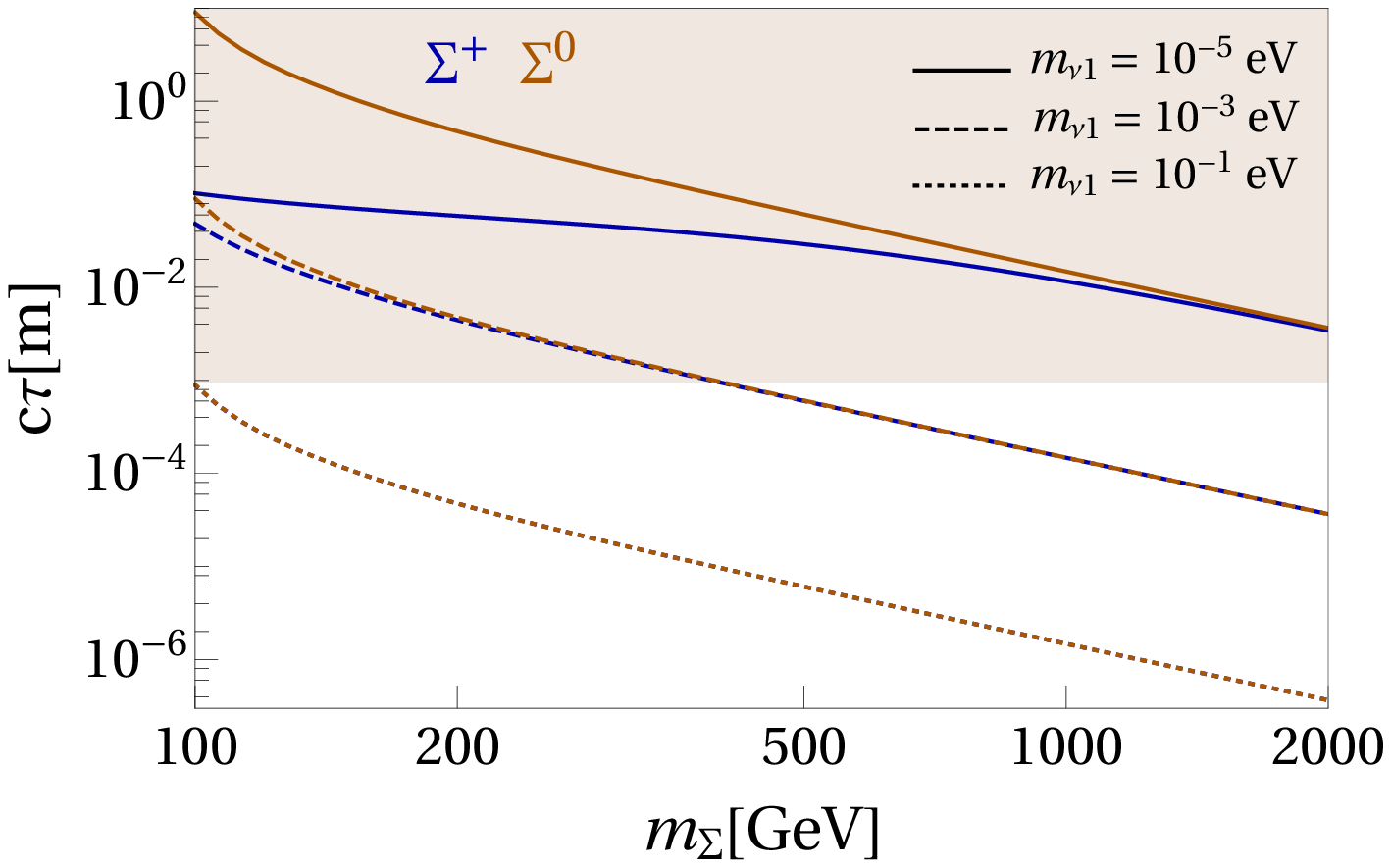}
\includegraphics[width=0.48\textwidth]{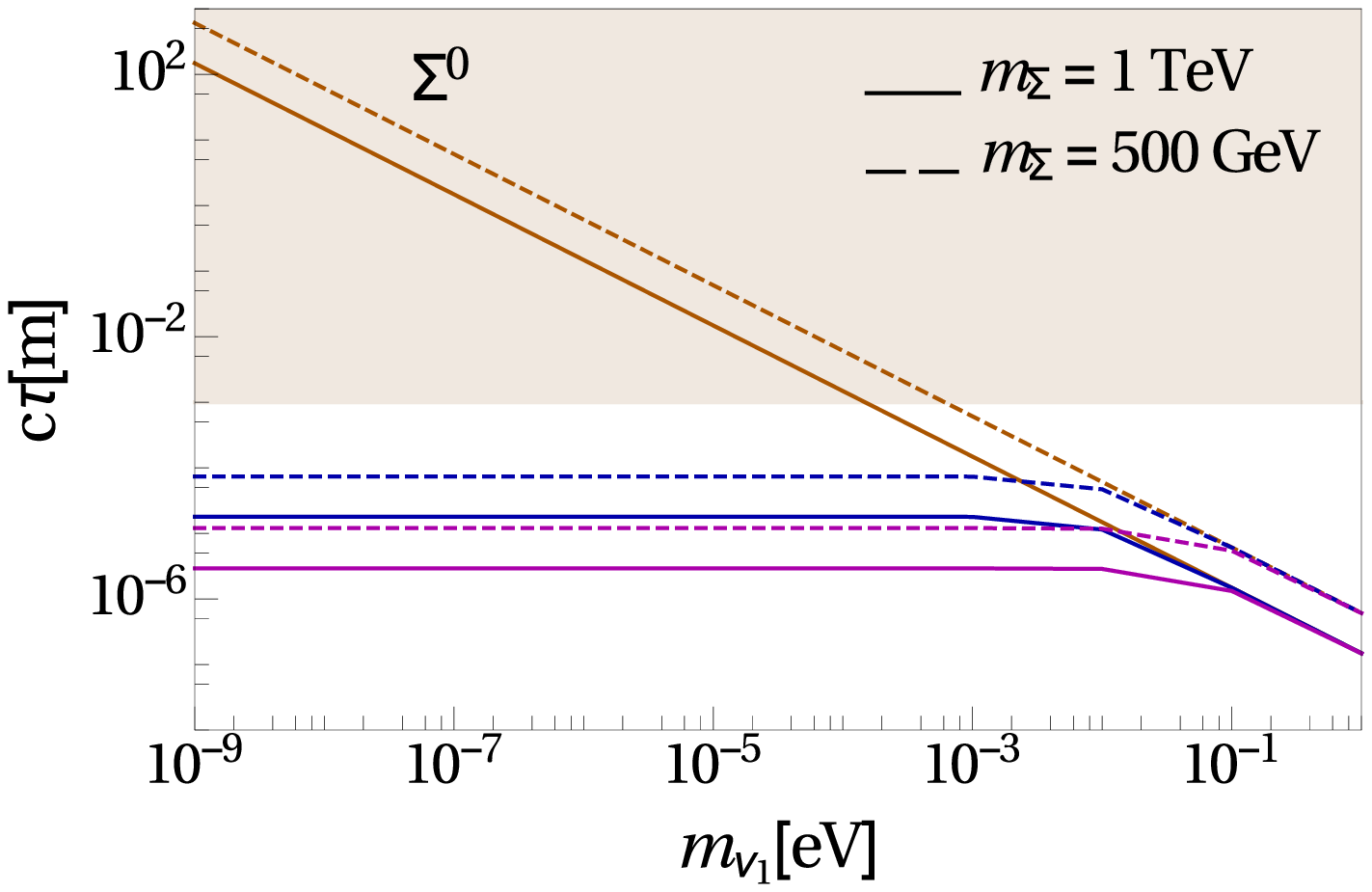} &
\end{tabular}
\caption{Left: Decay length for the neutral and charged components of
  the fermionic triplet $\Sigma$, as function of $m_{\Sigma}$, for
  several fixed choices of the lightest neutrino mass. The figure
  demonstrates that neutral and charged components of the fermionic
  triplet have nearly identical decay lengths. The gray band
  correspond to $c \tau > 1$ mm. Right: Decay length for the neutral
  component of the fermionic triplet versus the lightest neutrino mass
  $m_{\nu_{1}}$.  Solid (dashed) lines correspond to the case where
  $m_{\Sigma}= 1$ TeV ($m_{\Sigma}= 500$ GeV.)}
\label{fig:dwSST3}
\end{center}
\end{figure}

Example calculations for the decay lengths of the heavy fermions are
shown in fig. \ref{fig:dwSST3}.  Here and in all plots shown below,
unless noted otherwise, we fix the neutrino oscillation data at their
best fit point (b.f.p.) values, choose normal hierarchy and assume
that the matrix ${\cal R}$ is trivial. The plot to the left shows the
decay length, $c\tau$, for the neutral and charged components of the
fermionic triplet $\Sigma$, as function of $m_{\Sigma}$, for several
fixed choices of the lightest neutrino mass. Recall that $m_{\Sigma}$
below 390 GeV has already been ruled out by CMS
\cite{Sirunyan:2017qkz}.
%
%
Neutral and charged components of the fermionic triplet have nearly
identical decay lengths in the region of parameter space where $c\tau$
is smaller than roughly $c\tau \simeq 6$ cm. However, while $\Sigma^0$
can have much larger decay lengths, for $\Sigma^+$ there is an upper
limit on $c\tau$ of this order. This upper limit is due to the decays
$\Sigma^+ \to \Sigma^0 + \pi^+$.  Different from the decays to
standard model particles, the decays $\Sigma^+ \to \Sigma^0 + \pi^+$
are not suppressed by the smallness of the neutrino masses.  Instead,
because the masses of $\Sigma^+$ and $\Sigma^0$ are not exactly
degenerate, once 1-loop corrections are taken into account, there is a
small but non-zero decay width $\Gamma(\Sigma^+ \to \Sigma^0 + \pi^+)
\propto ( \Delta M)^3$, as pointed out first in
\cite{Franceschini:2008pz}.  Note that, $(\Delta M) \sim 160$ MeV, in
the limit $m_{\Sigma} \gg m_Z$, leading to a maximal $c\tau$ of order
6 cm for the charged state.\footnote{The maximal decay length is a
  function of $m_{\Sigma}$. For values below 400 GeV $c\tau$ could be
  (slightly) larger than 6 cm. However, recall that
  \cite{Sirunyan:2017qkz} puts a lower limit of $m_{\Sigma} \ge 390$
  GeV for prompt decays.}

The plot on the right shows the decay length for the neutral component
of the fermionic triplet versus the lightest neutrino mass
$m_{\nu_{1}}$. Here, we show two cases, fixing the mass of lightest
$\Sigma$ to $500$ GeV (dashed lines) and $1$ TeV (solid lines).
Orange, blue and magenta solid lines correspond to the cases where the
lightest of the three neutral triplets $\Sigma_{i}^0$ is the one
associated with $m_{\nu_i}$, $i=1,2,3$, where for normal hierarchy we
have $m_{\nu_{i}}=\{ m_{\nu_{1}},\sqrt{\Delta
  m_{\odot}^2+m_{\nu_{1}}^2}$, $\sqrt{\Delta m_{\rm
    Atm}^2+m_{\nu_{1}}^2} \}$. Blue and magenta lines are always in a
region with $c\tau$ below 1 mm, so one does not expect that at the LHC
a displaced vertex for these particles can be seen.  However, for the
case where the lightest $\Sigma$ is the one associated to $m_{\nu_1}$,
for small values of $m_{\nu_1}$ the decay length can be arbitrarily
large. This different behaviour is easily understood: Solar and
atmosperic neutrino mass squared differences require a minimum value
for $m_{\nu_2}\gsim 8.5$ meV and $m_{\nu_3}\gsim 50$ meV, while for
$m_{\nu_1}$ there is no lower limit experimentally.  For example, for
masses of $m_{\Sigma}$ to $500$ GeV (1 TeV), $c\tau \gtrsim 10^{-3}$
meter as long as the lightest neutrino mass is $m_{\nu_{1}} \lesssim
10^{-3}$ eV ($m_{\nu_{1}} \lesssim 2\times 10^{-4}$) eV.  For very
small values of $m_{\nu_{1}}$ the neutral fermions become quasi-stable
on LHC detector time scales, thus the displaced vertex signal
disappear (only missing energy is seen from $\Sigma^0$).  In this same
part of parametric space $\Sigma^+$ will have a decay length $c\tau
\sim 6$ cm, thus leaving a charged track signature inside the
detector.

\subsection{$d=7$}

We now turn to a discussion of the fermions in the BNT model. For this
model, the full neutral ($9\times 9$) and charged ($6\times 6$)
fermion mass matrices are given in eq. (\ref{eq:mzeroBNT}) and
(\ref{eq:mpBNT}). As shown in eq. (\ref{eq:mnuBNT}), for the BNT model
neutrino masses are essentially given by a linear seesaw. Here, there
are two independent Yukawa couplings, denoted by $Y_{\Psi}$ and
$Y_{\overline \Psi}$. This gives, in general, more parametric freedom
to fit neutrino oscillation data then can be described with the
standard Casas-Ibarra parametrization. A completely general
description for the neutrino mass fit for any Majorana neutrino mass
model has recently been given in
\cite{Cordero-Carrion:2018xre,Cordero-Carrion:2019qbp}.

However, here as everywhere else in this paper, we will be interested
again in the parameter region, where the decay length of the heavy
fermions is maximized.  To start the discussion, consider the mixing
matrices $U_{\nu}$ and $U_{l}$. For neutrinos one can estimate
$U_{\nu}^{H-L} \propto v(Y_{\Psi}\pm Y_{\overline \Psi}). M_{\bf 3}^{-1}$,
where $H-L$ indicates that we are considering here only the
off-diagonal parts of $U_{\nu}$, describing heavy-light mixing.
\footnote{There are two possible signs here, corresponding to the two
  components of the heavy quasi-Dirac neutrino.} $U_l^{H-L}$, on the
other hand, can be estimated by
$U_l^{H-L}\propto v Y_{\Psi}.M_{\bf 3}^{-1}$.  Neutral current
couplings, needed for decays such as $F_i \to f_j + Z^0$, are
proportional to $U_{\nu}$ ($U_l$) for neutral (singly charged)
fermions.  The total mixing matrix, $V$, entering the charged current,
on the other hand, is the product of $U_{\nu}.U_l^{\dagger}$, see
eq. (\ref{eq:dwII}), in case of $F^0$ and $F^+$. For charged current
decays of $F^{++}$ only $U_l$ enters. To identify the
parameter region, where maximal decay lengths occur, we then
have to find the minimal values for both, $U_{\nu}^{H-L}$ and
$U_l^{H-L}$.

Neutrino masses fix only the product of $Y_{\Psi}$ and
$Y_{\overline \Psi}$, i.e. the following simultaneous
scaling:
\begin{eqnarray}\label{eq:scale}
  Y_{\Psi}&\to& Y_{\Psi}^{'} = f Y_{\Psi} ,\\ \nonumber
  Y_{\overline \Psi}&\to& Y_{\overline \Psi}^{'}= (1/f) Y_{\overline \Psi},
\end{eqnarray}
leaves light neutrino masses unchanged for any value of $f$. At the
same time, it is easy to see that the ratio of
$U_l^{H-L}/U_{\nu}^{H-L}$ scales proportional to $f$, for $f\ll 1$.

\begin{figure}[tbh]
\begin{center}
\begin{tabular}{cc}
\includegraphics[width=0.45\textwidth]{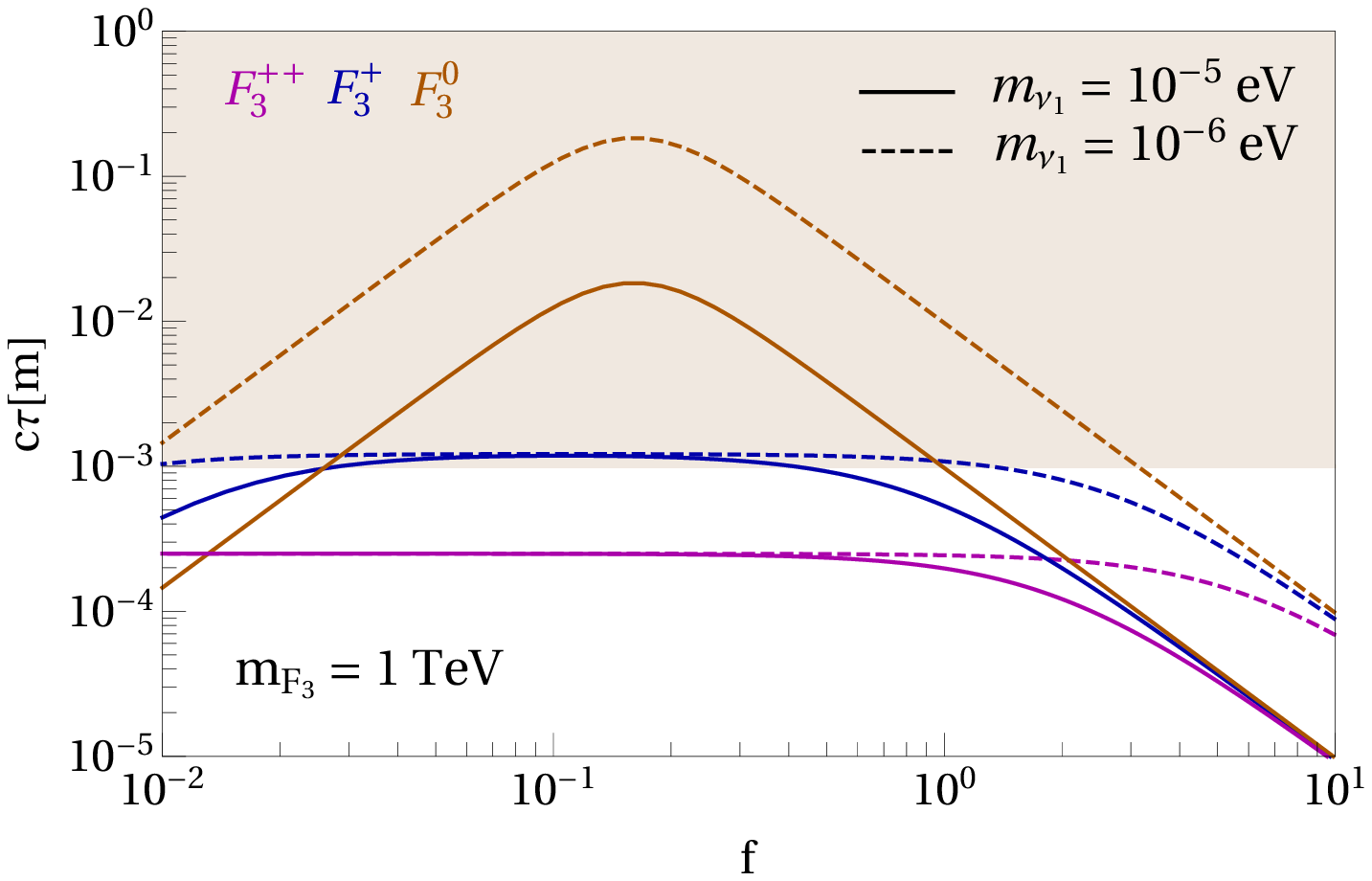} &
\includegraphics[width=0.45\textwidth]{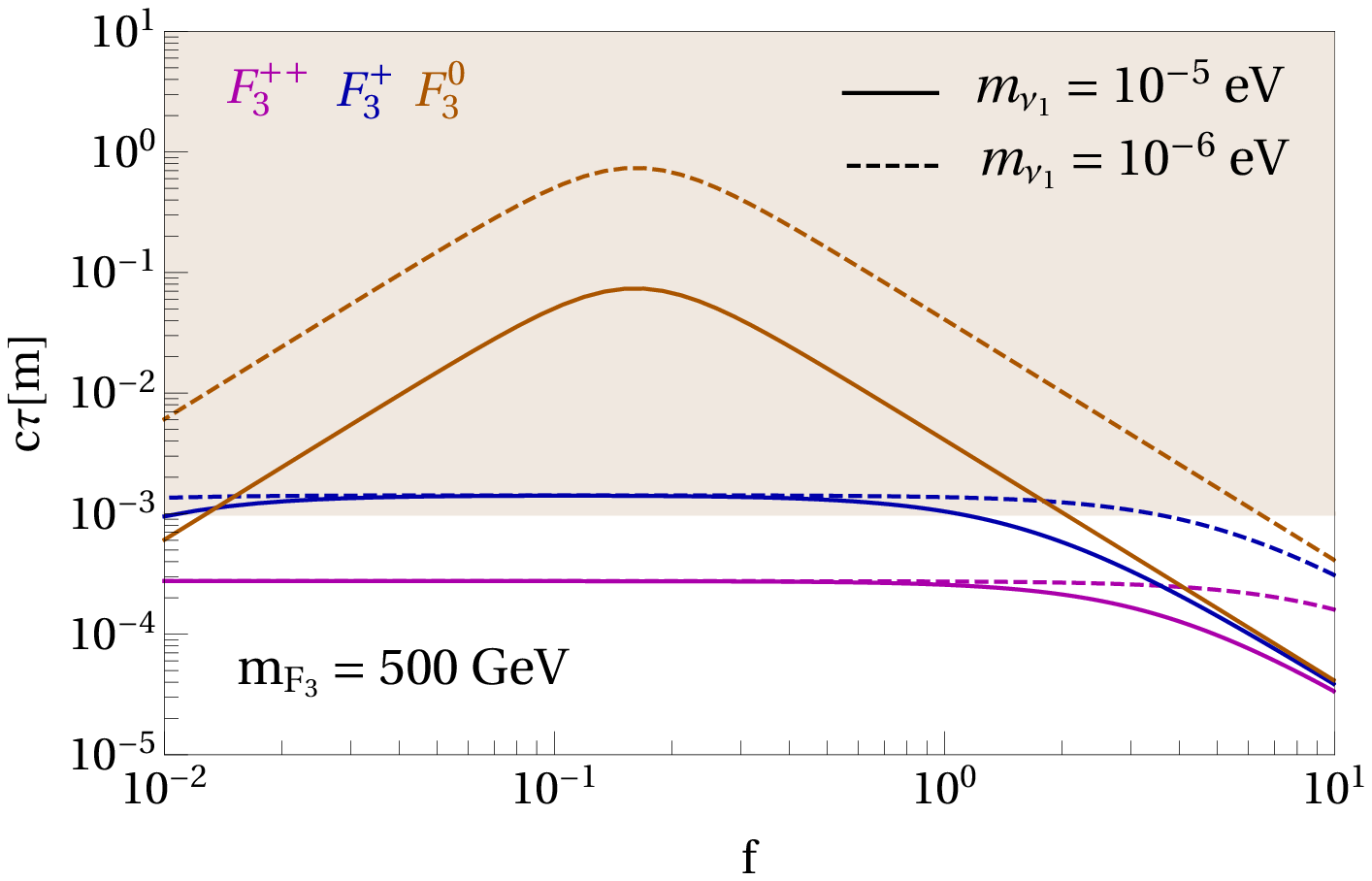} \\
\end{tabular}
\caption{Left: Decay lengths versus f for different values of the lightest neutrino mass and $m_{F_{3}} = 1$ TeV,  $v_{4}=2$ GeV. Right: Decay lengths versus f for different values of the lightest neutrino mass and $m_{F_{3}} = 500$ GeV,  $v_{4}=2$ GeV.}  
\label{fig:fScale}
\end{center}
\end{figure}

Fig. (\ref{fig:fScale}) shows some example decay lengths versus the
factor $f$ for different values of $m_{F_3}$ and the lightest neutrino
mass, $m_{\nu_1}$. Here, we have chosen $f=1$ to correspond to the
choice $Y_{\Psi}=Y_{\overline \Psi}$. Again, all other parameters have
been fixed to b.f.p values.
%
%
The figure demonstrates that for $f$ larger than roughly (1-5), the
decay lengths of $F^{++}$, $F^+$ and $F^0$ are similar and decrease
with increasing $f$. For smaller values of $f$ the decay length for
$F^0$ increases, until it reaches a maximum and then decreases again
for small values of $f$. This can be understood from the discussion
given above: Using small values of $f$ minimizes $U_l^{H-L}$, while
$U_{\nu}^{H-L}$ has a minimum near $f \simeq 1$ but increases for
both, large and small, values of $f$. Since both matrices enter into
the decays of $F^0$, the decay lengths of this particle are maximized
$f \simeq (0.1-0.3)$.

The maximal decay lengths for $F^+$ and $F^{++}$, however, are much
shorter. This again is due to decays to pions, i.e. $F^{++} \to F^{+}
+ \pi^+$ and $F^+ \to F^0 + \pi^+$. Similar to the case of the seesaw
type-III, discussed above, the mass splitting among the different
members of the multiplet are small, leading to small decay widths.
However, numerically $M(F^+)-M(F^0) \simeq 0.5$ GeV and
$M(F^{++})-M(F^+) \simeq 0.83$ GeV. Thus the maximal decay lengths
of $F^+$ and $F^{++}$ are smaller than the ones obtained for
$\Sigma^+$. 

\begin{figure}[tbh]
\begin{center}
\begin{tabular}{cc}
\hspace{-0.8cm}\includegraphics[width=0.46\textwidth, height=151px]{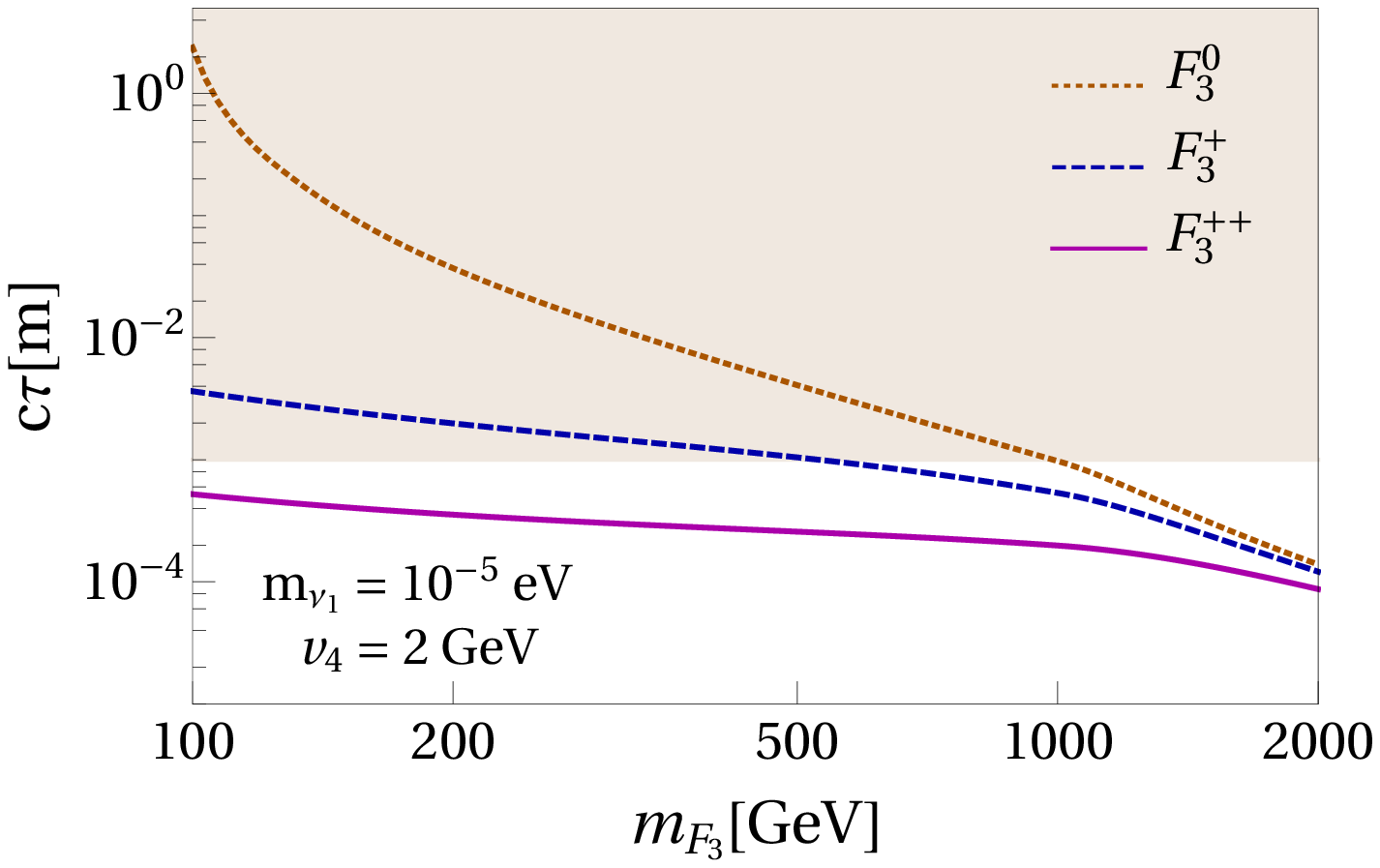} &
\includegraphics[width=0.45\textwidth, height=155px]{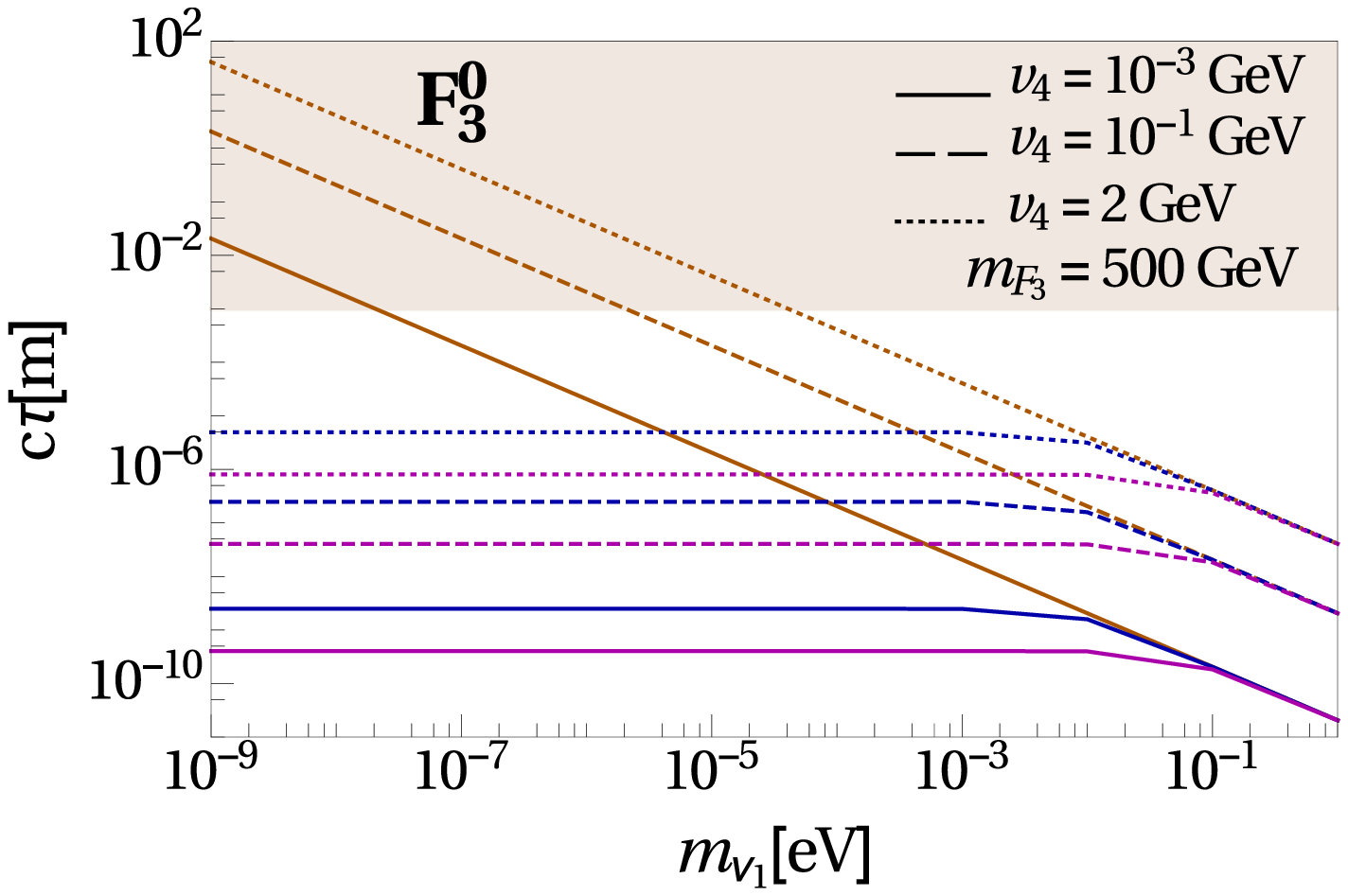} \\
\end{tabular}
\caption{Left: Decay length for the neutral, singly-charged and
  double-charged triplet components. In this case, the mass of the
  neutrino is fixed to $m_{\nu}=10^{-5}$ eV and the vev of the scalar
  quadruplet to $v_{\mathbf{4}}=2$ GeV. Right: Different comparisons
  for the $F_{3}^0$ decay lengths.  The dotted, dashed and continuous
  lines correspond to the choice of $v_{\mathbf{4}}$ to 2 GeV,
  $10^{-1}$ GeV and $10^{-3}$ GeV, respectively.  For both cases, the
  Yukawa matrices are fixed by the Casas-Ibarra parametrization, using
  the assumption $Y_{\Psi} \simeq Y_{\overline \Psi}$.}
\label{fig:fdFig2}
\end{center}
\end{figure}

In the rest of this subsection we will assume $Y_{\Psi}=
Y_{\overline \Psi}$.  In this part of the parameter space, the more
general neutrino fit of
\cite{Cordero-Carrion:2018xre,Cordero-Carrion:2019qbp} essentially
reduces to a Casas-Ibarra parametrization.
%
%
Choosing again b.f.p. values for the neutrino data and ${\cal R}=1$,
the decay length of the neutral, singly charged and doubly charged
components of ${\bf 3}_{1}^F= (F_{3}^{0},F_{3}^{+},F_{3}^{++}$) are
depicted in the left panel of Fig.\ref{fig:fdFig2}. The figure shows
that the maximal decay length for $F^+$ can be of the order of
$c\tau = 1$ mm, while $F^{++}$ decays always with lengths smaller
than mm. $F^0$, on the other hand, can have very large decay lengths.
The different behaviour of the widths of these states is again due
to the decays to pions, as discussed above. 

In the rest of this subsection, we will
discuss only the decay lengths of the neutral components. In the right
panel of Fig.\ref{fig:fdFig2} we show the decay length of $F_{3}^0$
for different values of the scalar quadruplet vev
$v_{\mathbf{4}}$. The solid, dashed and dotted lines correspond to the
cases where $v_{\mathbf{4}}=10^{-3}$ GeV, $v_{\mathbf{4}}=10^{-1}$ GeV
and $v_{\mathbf{4}}=2$ GeV respectively. For all the cases, we fixed
the mass of the $F_{3}^0$, denoted as $m_{F_{3}}$, to 500 GeV. Orange,
blue and magenta lines correspond again to the cases where the
lightest of the three neutral triplets $F_{3_i}^0$ is the one
associated with $m_{\nu_i}$, $i=1,2,3$. Larger values of the vev
$v_{\bf 4}$ imply smaller Yukawas, for the same values of light
neutrino masses, and thus larger lengths.  From the figure, we can
conclude that, for example, for a fixed value of $v_{\mathbf{4}} =
10^{-3}$ GeV, the decay length is $c\tau \gtrsim 1$ mm as long as
$m_{\nu_{1}} \lesssim 5 \times 10^{-7}$ eV, while for a fixed value of
$v_{\mathbf{4}} = 2$ GeV, the decay length is $c\tau \gtrsim 1$ mm as
long as $m_{\nu_{1}} \lesssim 5 \times 10^{-4}$ eV. Recall, that
there is an upper limit of roughly $v_{\bf 4} \lsim 3.5$ GeV, as
discussed in section \ref{sec:models}.

\subsection{$d=9$}

\begin{figure}
\begin{tabular}{cc}
\includegraphics[width=0.5\textwidth]{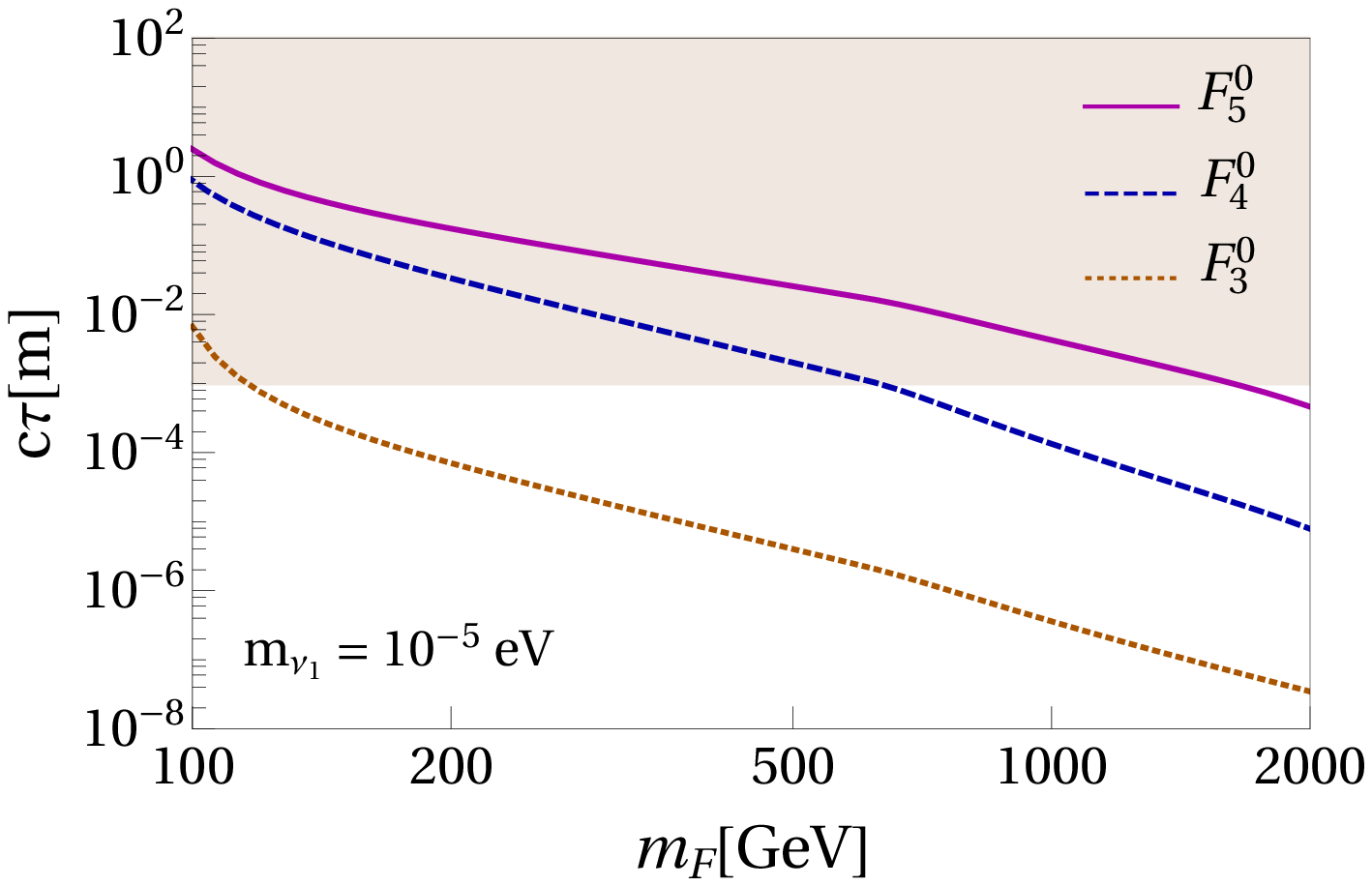} &
\includegraphics[width=0.5\textwidth]{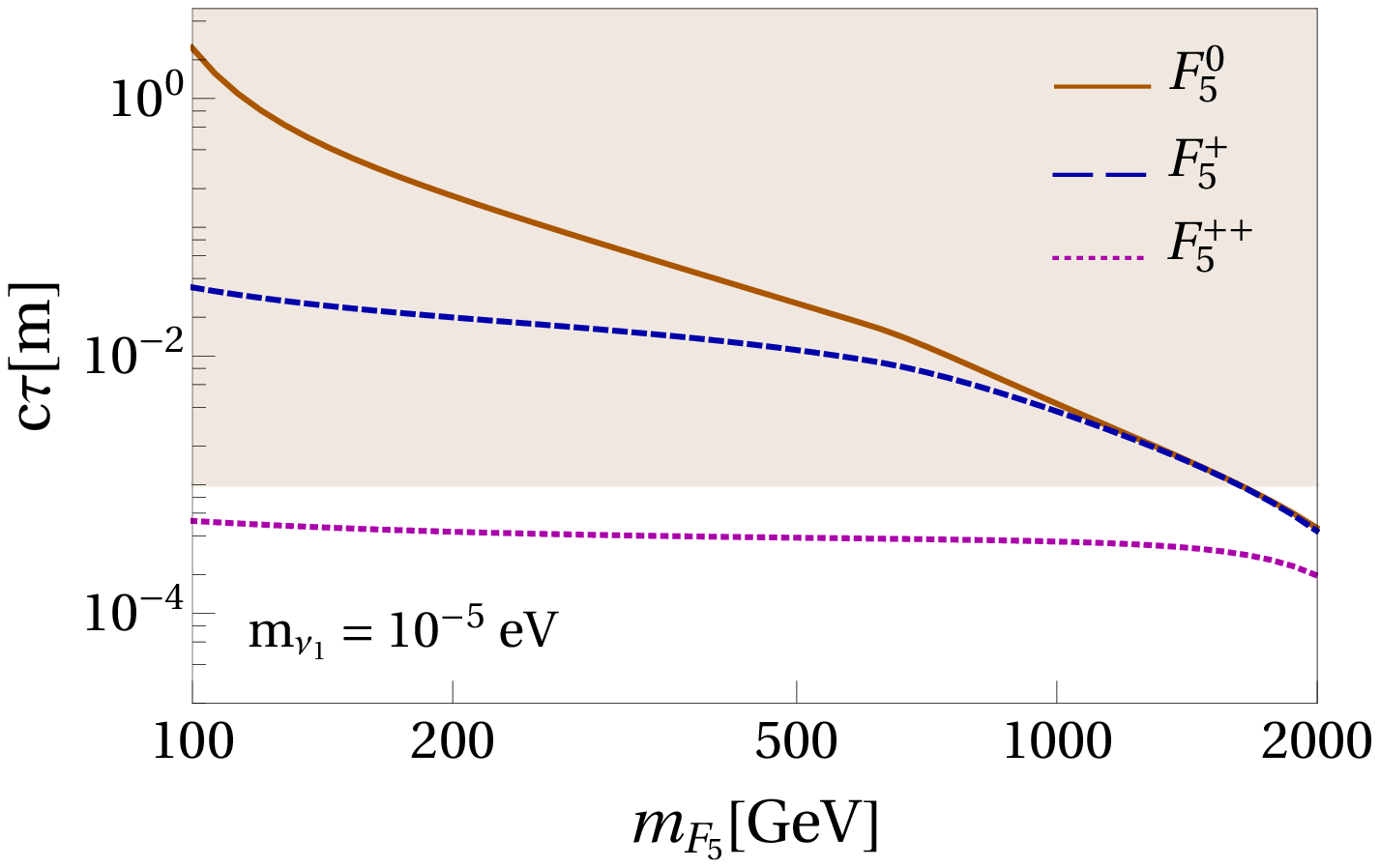} 
\end{tabular}
\caption{Left panel: Decay length for the neutral components of
  triplet, quadruplet and quintuplet fermion.  Right panel: Decay
  length for the neutral, singly-charged and double-charged components
  of a quintuplet. For both panels, the lightest neutrino is fixed to
  $m_{\nu}=10^{-5}$ eV and the values of the Yukawas
  ${\overline Y}_{\bf 34}= Y_{\bf 45}=1$ and ${\overline Y}_{\bf 45}=
  Y_{\bf 34}=0$.}
\label{fig:fdFig3}
\end{figure}

In the $d=9$ model, introduced in section \ref{sec:models}, there are
three different fermionic multiplets: $\mathbf{3_{1}^F}$,
$\mathbf{4_{1/2}^F}$ and $\mathbf{5_{0}^F}$. The decay lengths of the
lightest fermions then sensitively depends on which of these
multiplets is the lightest. This is shown in fig. (\ref{fig:fdFig3})
in the left panel.

Fig. (\ref{fig:fdFig3}), left panel, shows $c\tau$ as a function of
the mass of the fermion for a fixed choice of the lightest neutrino
mass. Here, we have fixed the diagonal elements of the Yukawas
${\overline Y}_{\bf 34}= Y_{\bf 45}=1$ and ${\overline Y}_{\bf 45}=
Y_{\bf 34}=0$. This choice of Yukawa couplings is motivated by the
fact that to lowest order neutrino masses do not depend on
${\overline Y}_{\bf 45}$ and $Y_{\bf 34}=0$, see section
\ref{sec:models}.  We set either $M_{\bf 3}$ or $M_{\bf 4}$ or
$M_{\bf 5}$ the smallest mass parameter, such that the lightest state
is (correspondingly) either mostly a triplet, a quadruplet or a
quintuplet fermion. While for a quintuplet (and partially also for a
quadruplet) fermion sizeable lengths appear, triplet fermions have
decay lengths two orders or more shorter than those found for the
quintuplet.

This fact can be understood as follows. The decay width of the
heavy fermions can be written roughly as:
\begin{equation}
    \Gamma(F_{3,4,5}^{0}) \propto |V_{3,4,5}|^{2}\frac{m_{F}^3}{m_{W}^2}.
\end{equation}
From the structure of the neutral fermion mass matrix in this model,
compare to eq. (\ref{eq:mnuDim9}), one can estimate very
approximately:
\begin{align}
    V_{3}&\simeq m_{D}M_{\bf 3}^{-1} \nonumber \\
     V_{4}&\simeq m_{D}M_{\bf3}^{-1}m_{34}M_{\bf 4}^{-1} \nonumber \\
     V_{5}&\simeq m_{D}M_{\bf 3}^{-1}m_{34}M_{\bf 4}^{-1}m_{45}M_{\bf 5}^{-1}
     \label{eq:vertex}
\end{align}
where $m_{D} \simeq \frac{v}{\sqrt{2}} Y_{\Psi}$,
$m_{34}=\frac{v}{\sqrt{6}}{\overline Y}_{\bf 34}$ and $m_{45}
=\frac{v}{2} Y_{\bf 45}$, respectively. Thus, one will find typically
$V_{5} < V_{4} \ll V_{3}$, if the $M_{\bf i}$ are larger than the
$m_{ij}$. Recall, that neutrino mass require at least some of these
parameter ratios to be small, see eq. (\ref{eq:numass1}).

%
%
In the right panel of fig. (\ref{fig:fdFig3}), we show the decay
lengths for the neutral, singly charged and doubly charged components
of $\mathbf{5_{0}^F}$. The figure demonstrates that the largest
lengths again are expected for the neutral component. The maximal
allowed decay length of $F_5^{+}$ ($F_5^{++}$) is around $c\tau \simeq
1.8$ cm ($c\tau \simeq 0.4$ mm) for masses larger than 400 GeV. The
reason is the same as discussed in the cases of seesaw type-III and
the BNT model: The charged components of the multiplet are slightly
heavier than the neutral one and, thus, can decay to the neutral state
plus a a charged pion with a width that is numerically small, but not
suppressed by the small neutrino mass.

\begin{figure}[tbh]
\begin{tabular}{ccc}
\hspace{-0.9cm}\includegraphics[width=0.35\textwidth, height=165px]{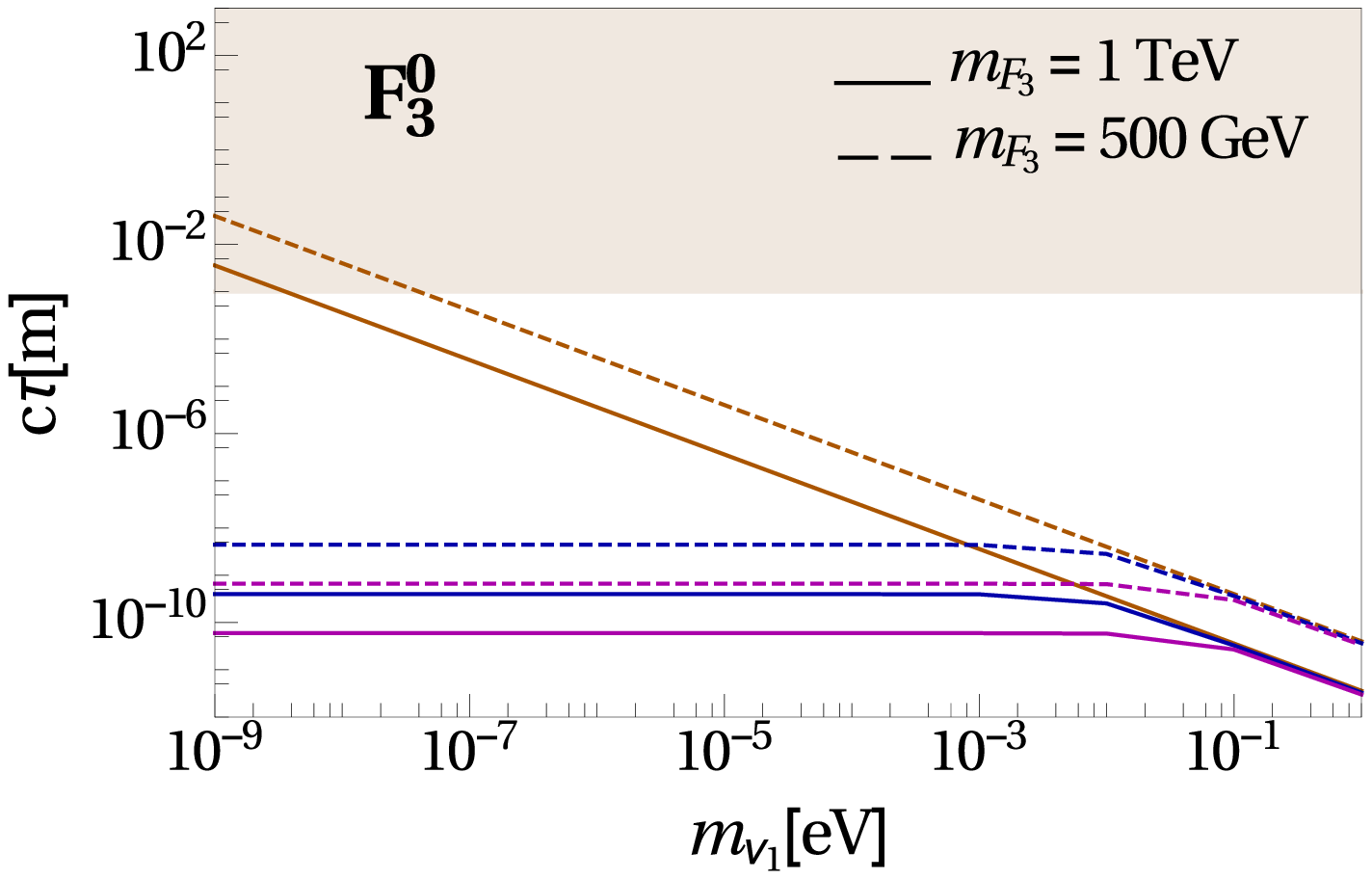} &
\includegraphics[width=0.35\textwidth, height=165px]{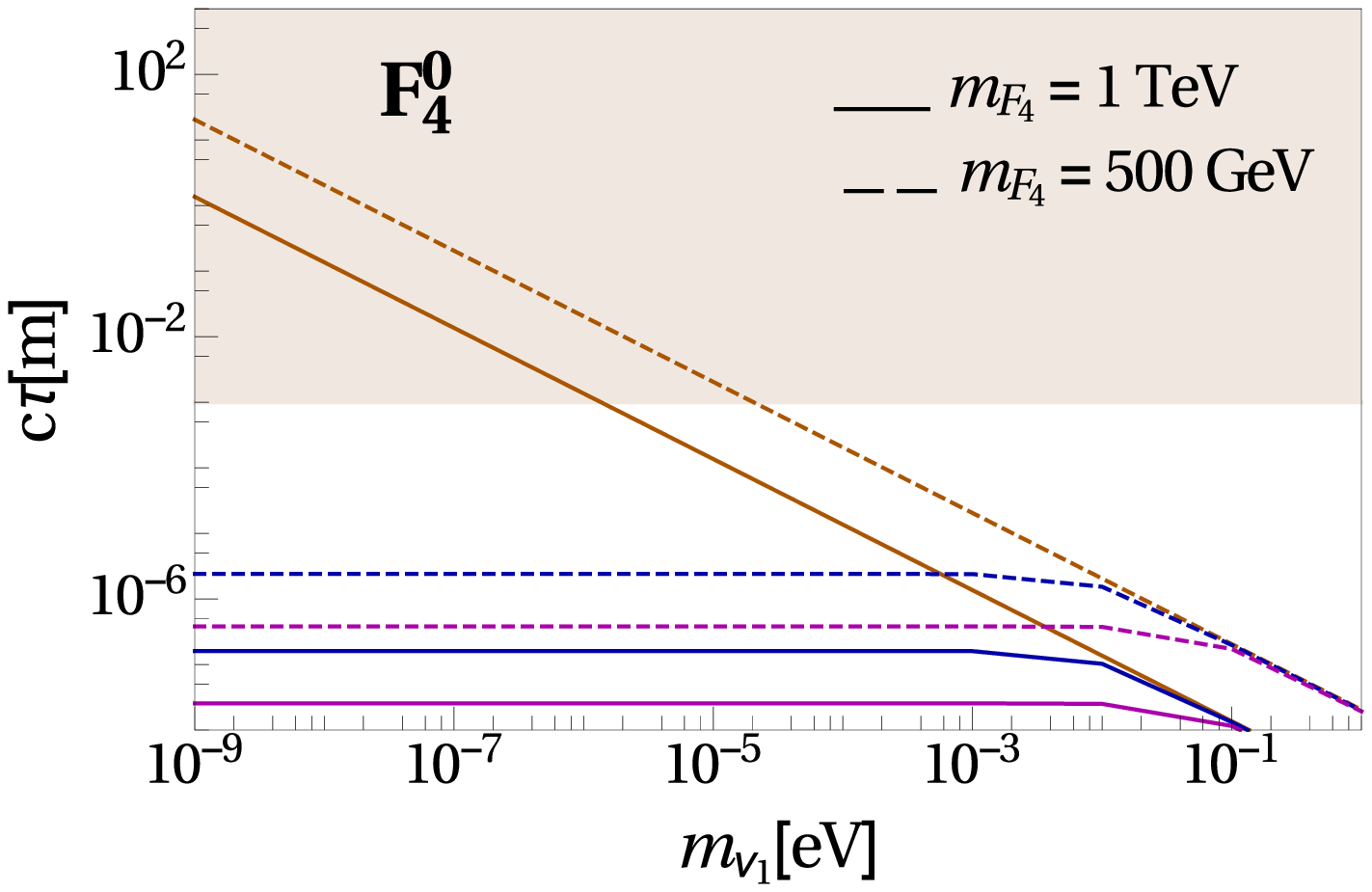} &
\includegraphics[width=0.35\textwidth, height=165px]{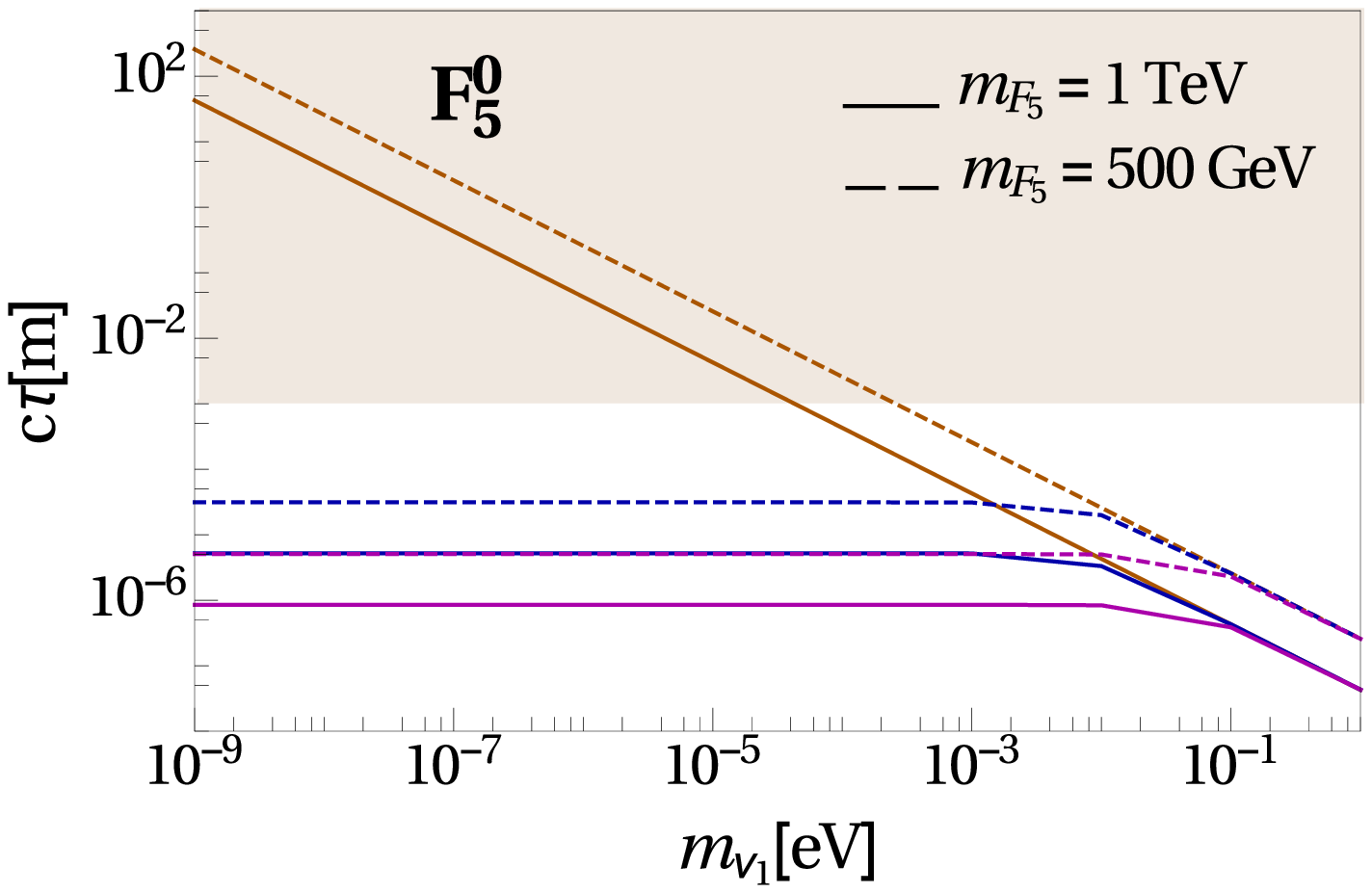} 
\end{tabular}
\caption{Decay lengths for the neutral component of the triplet (left
  panel), quadruplet (central panel) and quintuplet (right panel)
  fermion. For the three cases, the dashed and continuous lines
  correspond to the choice of the lightest fermion mass of 500 GeV and
  1 TeV respectively. Yukawas chosen as in fig. (\ref{fig:fdFig3}).}
\label{fig:fdFig4}
\end{figure}

The decay lengths of the neutral components of $\mathbf{3_{1}^F}$,
$\mathbf{4_{1/2}^F}$ and $\mathbf{5_{0}^F}$ are shown in
Fig.\ref{fig:fdFig4}. For the three panels, the orange, blue and
magenta lines again correspond to the cases where the lightest
fermions is associated to the first, second and third neutrino mass
eigenvalue, as obtained from a Cassa-Ibarra fit to neutrino data.  For
each case, the other free Yukawas of the model are fixed to 1, for
simplicity.

As expected, the region in parameter space where measurable decay
lengths occur are largest for fermions from the quintuplet, followed
by quadruplet fermions, while the triplet fermions have unmeasurably
short decay lengths practically in all acceptable parts of the
parameter space. Note the change in scale in the plot for the
triplets.

In summary, a systematic analysis of the fermionic decays from $d=5$
up to $d=9$ tree-level neutrino models allows us to conclude that for
all models larger decay lengths are found for a smaller overall
neutrino mass scale. In contrast larger masses of the fermions lead to
smaller widths in all cases.
%
%
We note that the charged components of the multiplets have
in all cases a maximal value for their decay lengths, imposed
by decays to the neutral fermion plus a charged pion. This decay
is suppressed by a small mass splitting, but not by the small
neutrino masses. We note that the maximal lengths for the
charged components are different in seesaw type-III, BNT and
the $d=9$ model. This might serve as a distinguishing feature
for the different models.

\section{Scalar decays} \label{sec:scalardesintegrations}

In this section we discuss the decays of the different scalars that
appear in the neutrino tree-level mass models at dimensions $d=5, 7$
and $d=11$, as introduced in section \ref{sec:models}. We will
concentrate on the main decay modes only and on identifying the
parameter region in which the decay lengths of the scalars are
maximized.

As discussed in the introduction, various experimental searches by the
LHC collaborations for type-II scalars (i.e. the case $d=5$) exist in
the literature, see for example
\cite{Aaboud:2017qph,CMS:2017pet,Aaboud:2018qcu}, to name the most
recent ones. The scalars of the $d=7$ and $d=11$ models can be
produced at the LHC in both, pair- and associated production. Since
pair production cross sections at LHC scale, at large masses,
proportional to the 4th power of the electric charge of the particle,
one expects in general larger cross sections at the LHC, and thus
better sensitivity to the scalars of these models than for seesaw
type-II. However, so far no dedicated LHC searches for these states
exist and, thus, there are no exact numbers on the lower limits on the
masses of these particles. We will, very roughly, assume that all
these scalars have masses in the range of [0.5,2] TeV, where the lower
end of the range could most likely already be excluded with current
data, if the final states of the scalar decays involve leptons.

\subsection{$d=5$}

The simplest case we consider corresponds to the type-II seesaw model
\cite{Schechter:1980gr,Mohapatra:1980yp, Cheng:1980qt} which has
already been widely studied in the literature \cite{Chun:2003ej,
  Perez:2008zc, Ferreira:2019qpf, Antusch:2018svb, Du:2018eaw,
  Dev:2018kpa, Dev:2018sel, Cai:2017mow, Bonilla:2017ekt,
  Reig:2016ewy, Mitra:2016wpr}. In particular the authors of
Refs. \cite{Du:2018eaw, Dev:2018kpa} studied the parameter region in
which the scalar mediators of the type-II seesaw model can be
relatively long-lived. Since this case can be understood as the
``proto-type'' for the scalar decays in our other models, we briefly
discuss seesaw type-II first.

As already mentioned, the type-II seesaw model adds a scalar SU(2)
triplet $\Delta = (\Delta^{0},\Delta^{+},\Delta^{++})$ to the SM.  We
mention that one expects the decay length of the scalars to have
similar values for different components of the triplet, especially for
large masses. Note, that this assumes that the different members of
the triplet have very similar masses, which is generally true for
large values of $m_{\Delta}$. We will therefore discuss only the
decays of $\Delta^{++}$.

The doubly charged scalar $\Delta^{++}$ has two decay modes:
$\Delta^{++}\rightarrow W^+ W^+$ and $\Delta^{++}\rightarrow l^+
l^+$.\footnote{Decay, such as $\Delta^{++}\to W^{+}+\Delta^{+}$ are
  usually kinematically not allowed in type-II seesaw for large
  $m_{\Delta}$. This is different in left-right symmetric extensions
  of the standard model with right-triplets \cite{Dev:2018kpa}, where
  the mass splitting between different components can be much larger
  than in the pure type-II seesaw case we consider here.}  These
partial decay widths can be expressed as \cite{Chun:2003ej,
  Perez:2008zc}:
\begin{equation}
\label{DW5}
\Gamma(\Delta^{++}\rightarrow  W^+ W^+) = \frac{\alpha_2}{32 } \frac{v^2_\Delta m^3_\Delta}{ v^2 m^2_W} I_2,  \ \ \ \ \Gamma(\Delta^{++}\rightarrow   l^+ l^+) = \frac{ 1}{16 \pi} \frac{ m_\Delta \sum_i m^2_{\nu_i} }{v^2_\Delta} 	
\end{equation}
with
\begin{equation}
I_2 = (1-4r_W+12r^2_W)(1-4r_W)^{1/2} \ \ ,\ \ r_W = (\frac{m_W}{m_\Delta})^2.
\end{equation}
Here, summation over the lepton generations $\alpha=e,\mu,\tau$ is
implicitly understood.  The leptonic decay channel is proportional to
the sum of the squares of the neutrino masses. This square has a
minimum value of roughly $\sum_i m^2_{\nu_i} \gsim (0.05
\text{eV})^2$, corresponding to a normal hierarchy of neutrino masses
with an atmospheric neutrino mass scale of $ m_{\nu_3} \simeq
\sqrt{\Delta m^2_{\rm Atm}}\simeq 0.05$ eV. Since
$\Gamma(\Delta^{++}\rightarrow W^+ W^+) \propto v^2_\Delta$, while
$\Gamma(\Delta^{++}\rightarrow l^+ l^+)\propto v^{-2}_\Delta$, there
exists a value of $ v_\Delta$ for which the total decay width
is minimized. The corresponding maximum for the decay length is
found for, in the limit $m_\Delta \gg m_W$:
\begin{equation}\label{eq:maxv}
  (v_{\Delta})_{max} = \Big(\frac{ 2 \sum_i m^2_{\nu_i} v^2 m^2_W}
                     { \alpha_2 m_\Delta^2 I_2}\Big)^{1/4}.
\end{equation}
At $v_\Delta$ smaller than this value, the decays are dominated by
leptonic final states, while for $v_\Delta$ larger than
eq. (\ref{eq:maxv}) gauge boson final states dominate. Interestingly,
the point in which the total decay width is minimized corresponds to
equal branching ratios of $\Delta^{++}$ decaying into leptons and
gauge bosons. For pair production $\Delta^{++}\Delta^{--}$ this
corresponds to final states $(l^+l^+l^-l^-:l^{\pm}l^{\pm}W^{\mp}W^{\mp}:
W^+W^+W^-W^-) \simeq (1:2:1)$, thus
maximizing the chances to see lepton number violation experimentally.

\begin{figure}[tbh]
\begin{center}
\includegraphics[width=0.57\textwidth,height=175px]{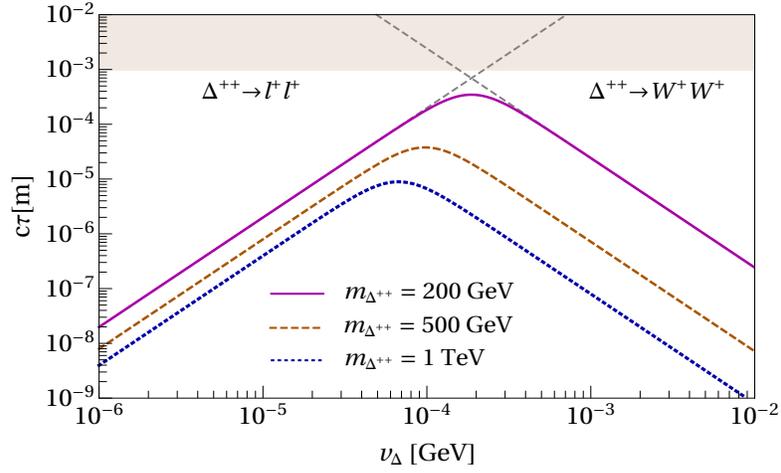}
\caption{Decay length ($c \tau$) versus $v_\Delta$. Three different
  choices for $m_{\Delta}$, corresponding to $m_{\Delta} = 200, 500,
  1000$ GeV are shown.  Dashed gray lines correspond to ``partial
  inverse widths'' $c\Gamma^{-1}(\Delta^{++}\rightarrow W^+ W^+)$ and
  $c\Gamma^{-1}(\Delta^{++}\rightarrow l^+ l^+)$ for $m_{\Delta} =
  200$ GeV. These lines are shown only for illustration.
  $\Delta^{++}\rightarrow W^+ W^+$ and $\Delta^{++}\rightarrow l^+
  l^+$ indicate the parameter range, where the corresponding final
  state dominates the decay width.}
\label{fig:d5}
\end{center}
\end{figure}

This can be seen also in Fig. \ref{fig:d5}. As shown in this plot, for
$m_\Delta = (1000, 500, 200)$ GeV there is maximum value of $c \tau$
as function $v_\Delta$. This maximum occurs at slightly smaller values
than predicted by eq. (\ref{eq:maxv}), $v_\Delta \sim (1, 2, 3)\times
10^{-4}$ GeV, due to phase space effects.  The dashed gray lines in
the background indicate the different partial widths for the case of
$m_{\Delta}=200$ GeV. These lines are for illustration purposes only,
to demonstrate where lepton or gauge boson final states are dominant.
The heavier (lighter) $\Delta^{++}$ the smaller (larger) the maximum
value of $c \tau $ becomes. Fig. \ref{fig:d5} also shows a gray band
which correspond $c\tau > 1$ mm. As we can see in this figure, given
current neutrino data, the scalar mediators of the seesaw type-II
always have $c \tau << 1$ mm for $m_\Delta > 200$ GeV. Thus, a doubly
charged scalar with $m > 200$ GeV, decaying with a visible decay
length can not give the correct explanation for the observed neutrino
masses, as expected from seesaw type-II.

\subsection{$d=7$}

We now turn to the case of the BNT model \cite{Babu:2009aq}.  The
scalar quadruplet with hypercharge $3/2$, can be written in components
as ${\bf 4^S_{3/2}} = (S^{3+}_4,S^{2+}_4,S^{+}_4,S^{0}_4)$.  Consider
first the triply charged scalar $S_4^{3+}$. It has two principal decay
modes: $S^{3+}_4\rightarrow W^+ W^+ W^+$ and $S^{3+}_4\rightarrow W^+
l^+ l^+$, where again we have suppressed flavour indices for the
leptons. Pair production of $S_4^{3+}$ can lead therefore to final
states with up to 6 $W$s or 4 leptons plus 2 $W$.  Of particular
theoretical interest are the LNV final states of $l^+l^+$+8 jets
(from hadronically decaying $W$s). 

In the limit where the mass of $S^{3+}_4$ is large, $m_4\gg m_W$,
one can find an approximate expression for the partial decay widths
of $S^{3+}_4$:
\begin{equation}
\label{eq:bnt}
\Gamma(S^{3+}\rightarrow W^+ W^+ W^+) \sim
\frac{3 g^6}{2048\pi^3} \frac{v^2_4 m^5_4}{m^6_W},
\hskip15mm
\Gamma(S^{3+}\rightarrow  W^+ l^+ l^+) \sim
\frac{ g^2}{3072\pi^3} \frac{ m^3_4 \sum_i m^2_{\nu_i} }{ v^2_4 m^2_W}.  
\end{equation}
Note that we have used here the phase space for massless final
state particles, thus eq. (\ref{eq:bnt}) approaches the numerical
result, see below, only for $m_4\gg m_W$.

Compared to the case of the seesaw type-II, discussed previously, one
notes that the partial widths are suppressed by phase space factors
for the three particle final state, but {\em enhanced} by different
additional factors $(m_4/m_W)^2$. The latter is due to the fact that
in the limit of large scalar masses the decays to the longitudinal
component of the $W$ dominates the total decay width. 

Most important, however, is that the decay width
$\Gamma(S^{3+}\rightarrow W^+ W^+ W^+)$ is proportional to $v_4^2$,
while $\Gamma(S^{3+}\rightarrow W^+ l^+ l^+)\propto v_4^{-2}$, similar
to the case of the seesaw type-II. Thus, using Eq. (\ref{eq:bnt}) we
can estimate a value $(v_4)_{max}$, which maximizes the decay length,
as:
\begin{equation}\label{eq:maxv4}
  (v_{4})_{max} \sim \frac{(\sum_i m^2_{\nu_i})^{1/4}m_W}{\sqrt{3}g\sqrt{m_4}}
\end{equation}
Roughly, this gives $(v_{4})_{max} \sim 2 \times 10^{-5}$ GeV for
$\sum_i m^2_{\nu_i}=(0.05)^2$ eV$^2$ and the example $m_4 \simeq 800$
GeV.

\begin{figure}[tbh]
\begin{center}
\includegraphics[width=0.57\textwidth, height=175px]{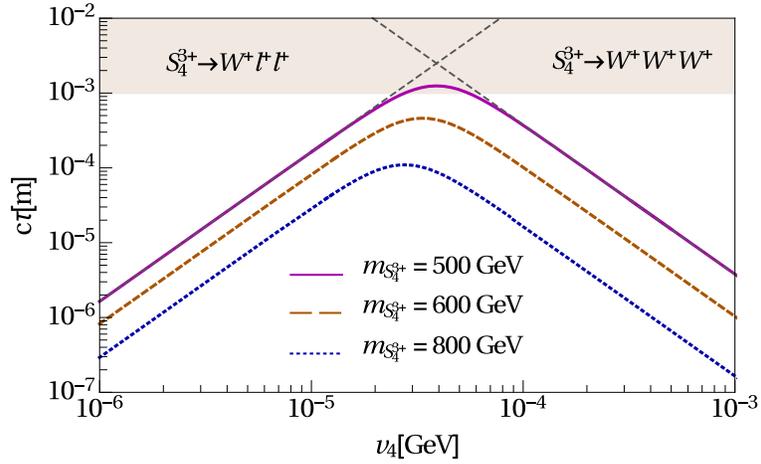} 
\caption{Decay lenghth ($c \tau$) versus $v_4$. Solid, dashed and
  dotted lines correspond to $m_{4} = 500, 600, 800$ GeV respectively.
  Dashed gray lines correspond to $c\Gamma^{-1}(S_4^{3+}\rightarrow
  W^+ W^+ W^+)$ and $c\Gamma^{-1}(S_4^{+3}\rightarrow W^+ l^+ l^+)$
  for $m_{4} = 500$ GeV.}
\label{fig:Fig5}
\end{center}
\end{figure}

With the \texttt{SARAH} generated model files for the BNT model, we
used \texttt{MadGraph}
\cite{Alwall:2007st,Alwall:2011uj,Alwall:2014hca} for a numerical
calculation of the partial widths of $S^{3+}_4$.  The results are
shown in fig. (\ref{fig:Fig5}). The figure shows $c\tau$ for
$S^{3+}_4$ for three values of $m_4$, $m_4 = 500, 600, 800$ GeV, as a
function of $v_4$. As before, the grey area indicates $c \tau > 1$ mm.
As discussed above, there is a value of $v_4$ for which the decay
length has a maximum, $(v_4)_{max} \sim (10^{-5}-10^{-4})$ GeV.
Comparison of the numerical values of $(v_4)_{max}$ with those
predicted by eq. (\ref{eq:maxv4}), show that the latter underestimates
the true value for small values of $m_4$, but is quite accurate for,
say, $m_4 \gsim 1$ TeV.  The figure also shows that the total decay
lengths, $c\tau$, should be shorter than $\sim 1$ mm for masses larger
than roughly $500$ GeV. Note that $c\tau$ does not include possible
boosts from production, such that visible decay lengths may occur
experimentally still for masses slightly larger than $500$ GeV.

We now turn to a brief discussion of the decays of  $S^{2+}_4$.
Similar to the case of the $\Delta^{++}$ of the seesaw type-II,
the main final states for the decays of the $S^{2+}_4$ are
$W^+W^+$ and $l^+l^+$. In the limit of large masses, $m_4\gg m_W$ 
we can estimate the ratio of the total widths as:
\begin{equation}\label{eq:rat23}
  \frac{\Gamma^{\rm tot}(S^{2+}_4)}{\Gamma^{\rm tot}(S^{3+}_4)}
  \sim \frac{1}{g^2}\frac{f(3)}{f(2)}\frac{m_W^2}{m_4^2}
  \simeq \frac{32\pi^2}{g^2}\frac{m_W^2}{m_4^2}
\end{equation}
Here $f(n) = 4 (4 \pi)^{2 n - 3} (n - 1)! (n - 2)! $ takes care of the
phase space volume available to the decay products of $n$ massless
particles, see ref. \cite{Fonseca:2018ehk}. From eq. (\ref{eq:rat23})
one estimates that $S^{2+}$ should decay with widths roughly a factor
$\sim 8$ ($\sim 5$) larger than $S^{3+}$ for masses order $m_4=800$
GeV (1 TeV). Comparison with the numerical results in
fig. \ref{fig:Fig5} shows that one can not expect to have a visible
decay length for $S^{2+}$, similar to what is shown in
fig. \ref{fig:d5} for the $\Delta^{++}$ of the seesaw type-II.

\subsection{$d=11$}

The last case we dissus are the scalars of the $d=11$ tree-level
neutrino mass model, defined in section \ref{sec:models}.  This model
introduces three different scalars: ${\bf 5}^S_2$, ${\bf 5}^S_1$ and
${\bf 3}^S_0$ as can be seen in Fig. \ref{fig:d9d11} (to the
right). Decay lengths of the scalars in this model depend on their
mass hierarchy. In our discussion we will assume the lightest heavy
particle is ${\bf 5}^S_2$, i.e. $m_{5_2} < m_{5_1}, m_{3}$. This is
motivated by the observation that the decay widths of the scalars to
final states with gauge bosons will scale with the square of the vev
of the neutral component of the corresponding scalar multiplet,
similar to the situation in the other models discussed above. 
As eqs (\ref{eq:tadv3})-(\ref{eq:tadv52}) show, from the tadpole
equations one expects that the vev of  ${\bf 5}^S_2$, $v_{52}$, is
numerically the smallest of the three vevs. Also,
${\bf 5}^S_2 = (S^{+4}_{5}, S^{+3}_{5}, S^{+2}_{5}, S^{+}_{5}, S^{0}_{5})$ 
contains a quadruply charged scalar, which can be expected to have
the smallest decay width of all the scalars in the model, due to
the additional phase space suppressions.

\begin{figure}[tbh]
\begin{center}
\includegraphics[width=0.57\textwidth, height=180px]{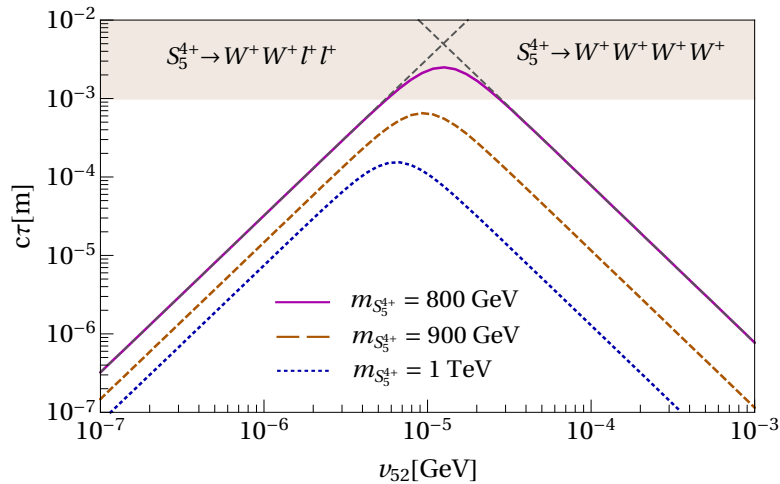} 
\caption{Decay lenghth ($c \tau$) versus $v_{52}$. Solid, dashed and
  dotted lines correspond to $m_{5_2} = 800, 900, 1000$ GeV
  respectively. }
\label{fig:Fig6}
\end{center}
\end{figure}

Fig. \ref{fig:Fig6} shows $c \tau$ for $S^{+4}_{5}$ versus $v_{52}$
for three different values of $m_{5_2}$, $m_{5_2} = 0.8, 0.9$ and $1$ TeV.
The two dominant decay modes here are $S_5^{4+}\rightarrow W^+ W^+ W^+ W^+$
and $S_5^{4+}\rightarrow W^+ W^+ l^+ l^+$. Again, the numerical
calculation of the widths were done using  \texttt{MadGraph}
\cite{Alwall:2007st,Alwall:2011uj,Alwall:2014hca}, based on 
our private version of the \texttt{SARAH} generated model files for
the $d=11$ model. 

As one can see in this figure there is a maximum around $v_{52} \sim
10^{-5} $ GeV, slightly different for different values of $m_{5_2} $.
The explanation for this feature is the same as in the seesaw type-II
and the BNT model, discussed previously. The leptonic decay mode
$S_5^{4+}\rightarrow W^+ W^+ l^+ l^+$ is proportional to $1/v_{52}^2$
while $S_5^{4+}\rightarrow W^+ W^+ W^+ W^+$ scales as
$v^2_{52}$. Therefore, there will be a maximum value $(v_{52})_{max}$
for which if $v_{52} << (v_{52})_{max} $ the leptonic mode will
dominate while if $v_{52} >> (v_{52})_{max} $ the decay mode $W^+ W^+
W^+ W^+$ will be the dominant contribution of $c \tau$.

As fig.  \ref{fig:Fig6} shows, there exists a region in parameter
space, for which $S_5^{4+}$ can decay with a visible decay length.
Visible decay lengths are possible up to masses very roughly in the
range of ($800-900$) GeV. The explanation for these larger masses
for the case of $S_5^{4+}$ compared to the $S_4^{3+}$ of the BNT
model is the large suppression from the 4-body phase space. This
suppression is only partially compensated by an additional factor
of $\frac{m_{5_2}^2}{m_W^2}$ in the decay width for the additional
$W$ boson in the final state.

We close the discussion of the $d=11$ scalars by mentioning that
all other scalars in the model should have a smaller decay length
than $S_5^{4+}$. The explanation for this is essentially smaller
phase space suppression for the decay widths of the other scalars,
very similar to the case of the BNT model discussed previously.

In summary, from the analysis of the scalar decays in tree-level
neutrino mass models at $d=5, 7, 11$ we can conclude that in all
cases there is an upper limit of the decay length, different from
the situation found for (neutral) fermions. For the scalars of the $d=7$ and
$d=11$ models this upper limit is of the order of a few millimeters,
and thus still in an observable range, while for the type-II seesaw
one does not expect to see a finite decay length experimentally. In
all cases this upper limit is reached in a region of parameter space
in which leptonic and gauge boson final states have similar branching
ratios.

\section{Conclusions}\label{sec:conclusions}

We have studied LHC phenomenology of different neutrino mass models.
All our models generate neutrino masses at tree-level. They range from
the simplest $d=5$ seesaws to $d=11$ models. Our main focus was to
study the decays of the heavy seesaw mediators. We concentrated on
masses below roughly 2 TeV, such that the heavy particles can be
produced at the high-luminosity LHC. We calculated the decay widths
for the fermions and scalars, that appear in the different models. Our
results depend on the unknown spectrum of mediators, i.e. we have to
distinguish two scenarios: (i) the exotic fermions are the lighter of
the mediators and (ii) scalars are lighter than the exotic fermions.

For case (i), fermions being the lighter (heavy) particles, we find
that their mixing with standard model particles is suppressed by the
light neutrino masses. Since there is no experimental lower limit on
the lightest neutrino mass, this mixing can be very small, leading to
very slow decays.  Very large decay lengths for neutral fermions are then
possible, even for fermions with masses order ${\cal O}$(TeV).  In
general, the fermions from models with $d>5$ tend to have smaller
decay lengths than those found for the fermions in $d=5$ models.

The situation is different for scalars. In high-dimensional models,
large scalar multiplets need to be introduced. Due to large phase
space suppression factors, the multiply charged scalars in these
models tend to have very small widths. As we have discussed scalars in
these models can decay either to pure gauge boson final states or to
final states with pairs of leptons (plus additional gauge
bosons). Since these two final states depend differently on the vacuum
expectation values of the scalars, there is always an upper limit on
the maximally allowed decay length. The numerical value of this
maximal decay length depends on the electric charge of the scalar
under consideration, but can not be larger than typically a few
millimeters. Interestingly the maximal decay length always appears in
a parameter region in which it should be possible to observe lepton
number violating final states.

Finally, as we have mentioned before, with the exception of the
well-known seesaw type-II, no LHC searches exist for any of the exotic
scalars we considered in this paper. Since these scalars can have
very high-multiplicity final states, we expect backgrounds to be low
at the LHC. This, combined with the large production cross sections
for multiply charged particles, makes LHC searches for these
exotic scalars very promising, in principle.

\bigskip

\centerline{\bf Acknowledgements}

\medskip
M. H. acknowledges funding by Spanish grants FPA2017-90566-REDC (Red
Consolider MultiDark), FPA2017-85216-P and SEV-2014-0398 (AEI/FEDER,
UE), as well as PROMETEO/2018/165 (Generalitat Valenciana). C.A. is
supported by Chile grant Fondecyt No. 11180722 and CONICYT PIA/Basal
FB0821. J. C. H. is supported by Chile grant Fondecyt No. 1161463.

\end{document}